\documentclass[12pt]{article}
\pdfoutput=1 
\usepackage{jheppub}
\usepackage{amssymb, amsmath}
\usepackage{graphicx}
\usepackage{hyperref}

\begin{document}
\begin{flushright}
INR-TH-2017-027
\end{flushright}
\vspace{-1cm}

\title{The fate of small classically stable $Q$--balls}
\author[a,b]{Dmitry Levkov}
\author[a,b]{Emin Nugaev}
\author[a,c,d]{Andrei Popescu}
\affiliation[a]{Institute for Nuclear Research of the Russian Academy
  of Sciences, 60th October Anniversary Prospect 7a,  Moscow 117312,
  Russia}
\affiliation[b]{Moscow Institute of Physics and Technology,
  Institutskii per.\ 9, Dolgoprudny 141700, Moscow Region, Russia}
\affiliation[c]{Faculty of Physics, Moscow State University, 
  Moscow 119991, Russia}
\affiliation[d]{Department of Applied Mathematics and Theoretical
  Physics, University of Cambridge, Wilberforce Road, Cambridge CB3
  0WA, UK}
\emailAdd{levkov@ms2.inr.ac.ru}
\emailAdd{emin@ms2.inr.ac.ru}
\emailAdd{popescu@ms2.inr.ac.ru}

\abstract{The smallest classically stable $Q$-balls are, in fact,
  generically metastable: in quantum theory they decay into
  free particles via collective tunneling. We derive
  general semiclassical method to calculate the rate of this
  process in the entire kinematical region of $Q$-ball
  metastability. Our method uses Euclidean field-theoretical solutions
  resembling the Coleman's bounce and fluctuations around them. As an
  application of the method, we numerically compute the decay rate 
  to the leading semiclassical order in a particular
  one-field model. We shortly discuss cosmological implications of
  metastable $Q$-balls.}
\maketitle

\section{Introduction}
\label{sec:intro}
It is well-known that $Q$-balls~\cite{Friedberg:1976me,
  Coleman:1985ki, Lee:1991ax}, the solitonic ``bags'' with attractive
globally charged particles inside, are rock stable if their total
charges $Q$ are large and the model has a mass gap $m$. Indeed, the
net masses $E_Q$ of these objects include negative energy of particle
attraction and, as a consequence, grow slowly as  the new particles
are added, ${E_Q \propto Q^{a}}$, $a<1$~\cite{Lee:1991ax}. This makes
sufficiently large $Q$-balls lighter than the sets of massive
particles with the same charge, 
$E_Q<mQ$, and their fission~--- forbidden by energy and charge
conservation.\footnote{If massless charged particles are present, the
  $Q$-balls may evaporate by emitting them from the
  surface~\cite{Cohen:1986ct, Multamaki:1999an}. Below 
  we assume that this process, if exists, occurs at negligibly small
  rate.}

Exceptional stability singles out $Q$-balls as the favorite toys for
cosmologists~\cite{Gorbunov:2011zz}. These objects may be generated in
the early Universe via first-order phase
transition~\cite{Frieman:1988ut, Griest:1989cb,   Krylov:2013qe} or
fragmentation of  the scalar condensate~\cite{Kusenko:1997si,
  Kasuya:2014bxa, Zhou:2015yfa} to form dark matter at the present
epoch. They may also participate in the  Affleck-Dine
baryogenesis~\cite{Affleck:1984fy,Dine:2003ax,   Allahverdi:2012ju}
catalyzing the process and hiding the baryon number from the sphaleron
transitions~\cite{Enqvist:1997si,   Enqvist:1998en}. In the recklessly
exotic scenario $Q$-balls may even impersonate black
holes~\cite{Troitsky:2015mda} posing new challenges for the
astrophysical observations~\cite{Falcke:1999pj, Kardashev:2015xua}. As
a pleasant bonus, the objects of this kind invariably
appear~\cite{Enqvist:2003gh} in supersymmetric models with flat
directions which remain in the club of our favorite extensions of the
Standard Model despite obvious lack of support from the experimental  
data~\cite{Giordano:2017dqe}.

\begin{figure}[htb]
  \hspace{3cm}(a) \hspace{3.6cm}(b) \hspace{4.8cm}(c)

  \vspace{-3mm}
  \hspace{-1mm}\includegraphics[trim=0 5 0 0]{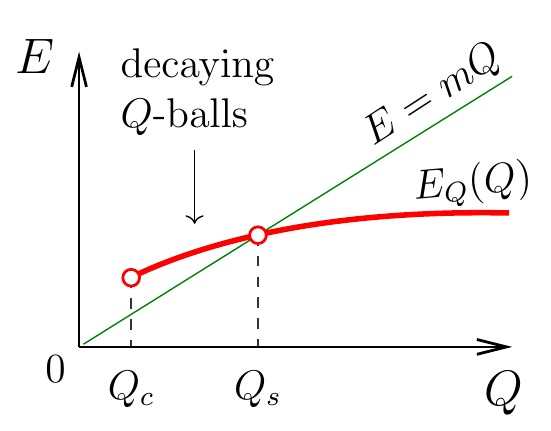}\hspace{5mm}
  \includegraphics{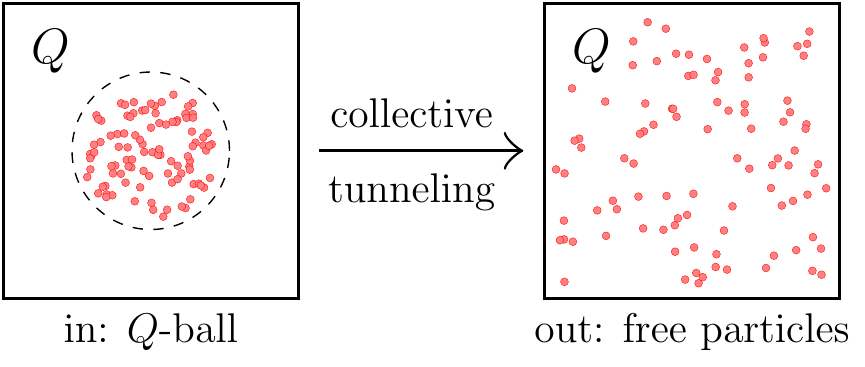}
  \caption{(a) Typical dependence of $Q$-ball mass $E_Q$ on its
    charge $Q$ (thick solid line). (b),~(c)~Decay of small $Q$-ball
    into free particles. \label{fig:intro}}
\end{figure}

Weak dependence of $Q$-ball mass $E_Q$ on its charge has another, less
widely recognized consequence. At smallest $Q$ this mass exceeds the
minimal energy $mQ$ of free particles, which makes small   $Q$-balls
generically unstable~\cite{Friedberg:1976me, Coleman:1985ki, 
  Lee:1991ax} in some region ${Q<Q_s}$, $E_{Q_s} = mQ_s$,
see Fig.~\ref{fig:intro}a. Still, these objects are 
classically stable if $Q$ is slightly below $Q_s$. Indeed, the energy
deficit $E_Q-mQ$ of their decay would be small, as well as the momenta
of the final-state particles. But finite-size solitons like that
cannot evolve classically into waves/particles of much larger
wavelength! This gives a range of charges $Q_c < Q < Q_s$ where the
$Q$-balls exist, remain classically  stable and  yet, decay into free
particles quantum-mechanically. Being local
minima of energy at fixed charge, they cannot ``leak'' the particles
one-by-one. Rather, they remain metastable for a long time and then
explode  in a spectacular collective tunneling event sketched in
Fig.~\ref{fig:intro}b. We stress that the metastability window  $Q_c <
Q < Q_s$ of global $Q$-balls appears in all three-dimensional models
with bounded energy considered in literature, see
e.g.~\cite{Friedberg:1976me, Lee:1991ax, Gulamov:2013ema,
  Nugaev:2013poa}.

In this paper we develop general semiclassical method to compute 
the decay rate of  metastable $Q$-balls at $Q,\, Q_c,\, Q_s\gg 1$. To
the best of our knowledge, no method of this kind has been ever proposed
before, with the closest example describing quantum collapse of a
metastable Bose condensate in condensed matter
physics~\cite{Freire:1999}. There are two reasons, why. First, the
states of $Q$-balls are not just the local minima of energy like the
false vacua~\cite{Kobzarev:1974cp}, but  the minima at a fixed charge
$Q$. Second, the charged scalar particles inside the $Q$-ball are
typically described by an oscillating complex scalar field
$\varphi(\boldsymbol{x},\, t) \propto  \mathrm{e}^{i\omega  t}$ which  
grows without bound in Euclidean time ${\tau \equiv it}$. These features do
not enable one to use powerful techniques developed for false vacuum
decay~\cite{Coleman:1977py, Callan:1977pt, Coleman:1978ae}, or even 
perform Wick rotation at the start of calculations.   

Our semiclassical method is derived by evaluating path
integral for the $Q$-ball decay rate in the saddle-point
approximation~\cite{Khlebnikov:1991th, Demidov:2011dk,
  Demidov:2015bua, Demidov:2015nea, Andreassen:2016cvx} and  going into
Euclidean time 
afterwards. This general procedure is justified at $Q\gg 1$. We arrive  
at the semiclassical expression for the rate in terms of a
certain classical 
solution $\varphi = \varphi_{cl}(\boldsymbol{x},\, \tau)$ which
resembles the Coleman's  bounce~\cite{Coleman:1977py} in Euclidean
time $\tau$ and yet,  
differs from it in important details. Our solution interpolates between the
$Q$-ball at $\tau\to -\infty$ and configuration in the catchment area
of the vacuum at $\tau  =0$. Unlike the bounce, it
can be continued to Euclidean time only in a very specific way, by 
turning the charged field $\varphi$ and its complex conjugate
$\bar{\varphi}$ into independent real functions of $\boldsymbol{x}$
and $\tau$. The real ratio  $\varphi_{cl}/\bar{\varphi}_{cl}$ 
satisfies  certain boundary conditions at $\tau = 
-\infty$ and $\tau=0$. Given  the solution, one computes the
probability of $Q$-ball decay per unit time, 
\begin{equation}
  \label{eq:8}
  \Gamma_Q = A_Q \cdot\mathrm{e}^{-F_Q}\;,
\end{equation}
where the suppression exponent $F_Q$ is related to Euclidean action on
$\varphi_{cl}$ and the prefactor $A_Q$ includes fluctuation
\begin{sloppy}
determinant around this solution. In the semiclassical limit $Q \sim
Q_c\gg 1$ the exponent ${F_Q \propto Q_c}$ becomes large, and ${A_Q
\sim m Q_c^{1/2}}$.  

\end{sloppy}

As a by-product, we obtain expression for the width of metastable
$Q$-balls interacting with neutral finite-temperature bath.  In this case
the semiclassical solution lives on a finite Euclidean time interval
and expressions for the exponent and the prefactor are modified
accordingly.

\begin{figure}[htb]
  \vspace{3mm}
  \hspace{3.5cm} (a) \hspace{6.4cm} (b)

  \vspace{-3mm}
  \centerline{
    \includegraphics[trim=0 -5 0 0]{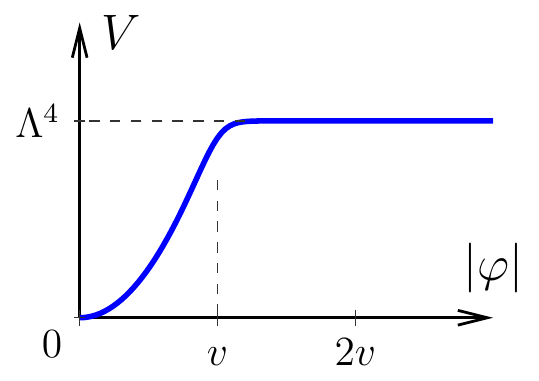}\hspace{8mm}
    \includegraphics[trim=0  5 0 0]{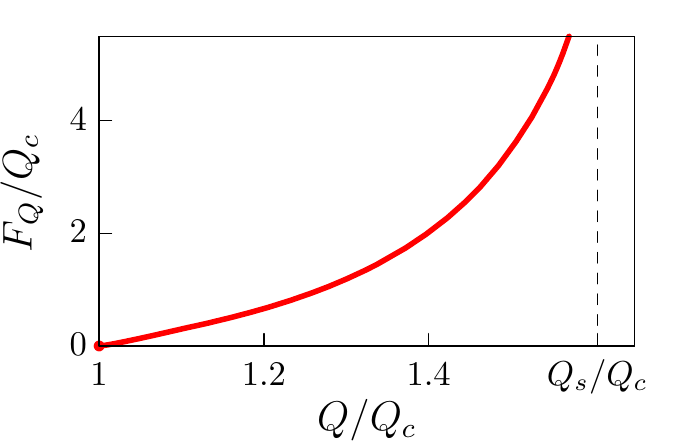}}

  \caption{(a) Scalar potential used in the numerical
    calculations. (b)~Suppression exponent for the rate of  $Q$-ball decay
    in the model with potential (a).\label{fig:result}}
\end{figure}

We illustrate the semiclassical method in the model of a complex
scalar field~$\varphi(\boldsymbol{x}, \, t)$,
\begin{equation}
  \label{eq:13}
  S = \int d^3 \boldsymbol{x} dt\left[ \partial_\mu \varphi \partial^\mu
    \bar{\varphi} - V(\varphi \bar{\varphi}) \right] \;,
\end{equation}
with scalar potential shown in Fig.~\ref{fig:result}a. Overbar in
Eq.~(\ref{eq:13}) and below denotes complex conjugation, expression
for $V(\varphi\bar{\varphi})$ will be given in the main
text. Note that our model  approximates renormalizable
Friedberg-Lee-Sirlin two-field  model~\cite{Friedberg:1976me} in the
limit when the additional field is very massive and can be
integrated out. Besides, the potential in Fig.~\ref{fig:result}a is
flat at large fields: $V\approx \Lambda^4$ at $|\varphi|>v$.  In this
regard it is similar to potentials appearing in supersymmetric
theories with flat  directions~\cite{Enqvist:2003gh}.

The model (\ref{eq:13}) has a family of $Q$-ball solutions  with
mass graph $E_Q = E_Q(Q)$ similar to the one in
Fig.~\ref{fig:intro}a. These objects are metastable at charges between
${Q_c   \approx 266 \, v^4/\Lambda^4}$ and $Q_s \approx 1.6 \,
Q_c$. At $v > 5 \, \Lambda$ the values of $Q_c$ and $Q_s$ are large
and semiclassical approximation can be used for all metastable
$Q$-balls.  We numerically found the respective semiclassical
solutions  $\varphi_{cl}(\boldsymbol{x},\, \tau)$ and computed the
suppression exponent $F_Q$ of the decay rate, see
Fig.~\ref{fig:result}b. As expected, the value of $F_Q$ tends to zero
and infinity in the limits $Q\to Q_c$ and $Q\to Q_s$ corresponding to 
classically decaying and absolutely stable  $Q$-balls. Leaving
numerical calculation  of the fluctuation determinant to further
studies, we roughly estimate the prefactor in Eq.~(\ref{eq:8}) as
${A_Q \sim mQ_c^{1/2} \sim 10\, v}$. 

In cosmology metastable $Q$-balls may form naturally long-living
decaying dark matter. In the most exciting scenario the lifetime of
these objects is comparable to the age of the Universe, so that their
decays can affect structure formation and 
alleviate~\cite{Chudaykin:2016yfk} emerging tension between the
low-redshift~\cite{Riess:2011yx, Freedman:2012ny} and high-redshift
(CMB)~\cite{Ade:2015xua} cosmological measurements. This requires,
however,  moderately small $Q$-ball charges $Q\sim Q_c \sim F_Q$: in
the above model $\Gamma_Q^{-1}$ is of order $10^{10}$~years
if ${Q\sim Q_c\sim 10^{2}}$, see Fig.~\ref{fig:result}b. Our estimates
show that the $Q$-balls with these charges may be generated in the
early Universe via phase transition or fragmentation of the scalar
condensate if the generation temperature is raised to $T\sim 10^{11}
\, \mbox{GeV}$, cf.~\cite{Kusenko:2001vu}. Besides, alternative
generation mechanisms~--- say, pair production~\cite{Kusenko:1997ad}
---  may be essential at small $Q$. This suggests a path to new
cosmological scenarios with metastable $Q$-balls which may be
considered separately.

In the introductory Sec.~\ref{sec:semicl-meth-its} we explain 
our semiclassical method and show main numerical
results. Derivation of  the method and details of its 
numerical implementation are given in
Secs.~\ref{sec:deriv-semicl-meth} and \ref{sec:numer-impl}, 
respectively. We discuss possible cosmological applications of the
metastable $Q$-balls in Sec.~\ref{sec:gener-prosp}. 
\section{Semiclassical method and its application}
\label{sec:semicl-meth-its}

\subsection{Metastable $Q$-balls}
\label{sec:small-q-balls}
We start by briefly reviewing the properties of small $Q$-balls. To be
concrete, consider the model~(\ref{eq:13}) of complex scalar
field $\varphi$ with the potential 
\begin{equation}
  \label{eq:2}
  V = - \frac{m^2v^2}{b}\log\left(\frac{\mathrm{e}^{-b\varphi
      \bar{\varphi}/v^2} + \mathrm{e}^{-b} }{1+
    \mathrm{e}^{-b}}\right) \;, \qquad b=8
\end{equation}
shown in Fig.~\ref{fig:result}a. The scalar bosons in this model have 
mass $\approx m$ in vacuum $\varphi=0$ and  become almost
massless at large $\varphi$: $V \to \Lambda^4 \equiv m^2 v^2
\log(\mathrm{e}^b+1) /b$ as ${|\varphi| \to +\infty}$. This corresponds
to short-range attraction between the bosons. Importantly, the
potential (\ref{eq:2}) approximately describes decoupling limit of the
renormalizable  Friedberg-Lee-Sirlin model~\cite{Friedberg:1976me}, as
we will argue  in Sec.~\ref{sec:gener-prosp}.

\begin{figure}[htb]
  \hspace{4.4cm} (a) \hspace{6.0cm}(b)
  
  \centerline{\includegraphics[trim=0 4 0 10]{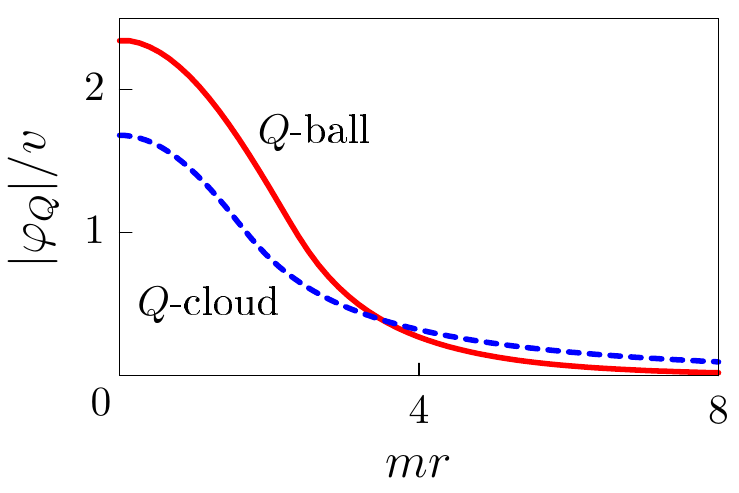}
  \includegraphics[trim=0 -7 0 2, scale=0.92]{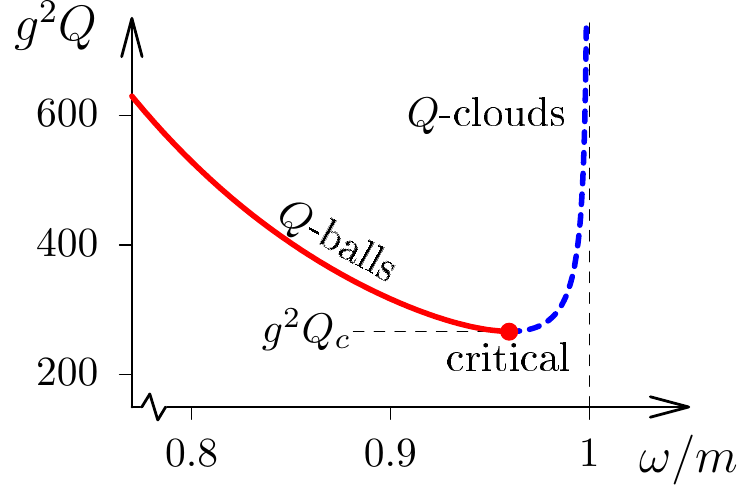}}

  \caption{(a)~Profiles $|\varphi_Q|$ of 
    nontopological solitons with $\omega \approx 0.88\, m$ and $0.99\,
    m$ in the model (\ref{eq:2}) (solid and dashed lines,
    respectively). The charges of these solutions are equal:
    $g^2 Q = 353$ (${Q/Q_c \approx  1.33}$). (b) Soliton
    charge $Q$ as a function of its frequency~$\omega$. \label{fig:Qw}}
\end{figure}

The model~(\ref{eq:13}) possesses global conserved
charge
\begin{equation}   
  \label{eq:1}
  Q = i \int d^3\boldsymbol{x} \left ( \varphi \partial_t
  \bar{\varphi} - \bar{\varphi} \partial_t \varphi \right)\;,
\end{equation}
related to phase rotation symmetry $\varphi \to
\mathrm{e}^{-i\alpha}\, \varphi$, $\bar{\varphi} \to
\mathrm{e}^{i\alpha} \bar{\varphi}$. Conservation of this
quantity is vital~\cite{Coleman:1985ki} for the existence and
stability of $Q$-balls~--- nontopological solitons carrying nonzero
$Q$.
\begin{sloppy}

In what follows we use semiclassical approximation which is
justified at ${g \equiv m/v \ll 1}$. Indeed, change  of
variables $\varphi \to v\varphi$, $x^\mu \to x^\mu/m$ eliminates all
parameters from the classical action~(\ref{eq:13}), (\ref{eq:2})
except for the combination $g^{-2}$ right in front of it. This means
that $g^2 \ll 1$ is the semiclassical  parameter in our
model. Besides, the charge (\ref{eq:1}) also becomes proportional to
$g^{-2}$ after the change of variables suggesting that $Q\gg 1$ in the
semiclassical regime.

\end{sloppy}
The simplest way to find nontopological solitons in the
model~(\ref{eq:2}) is to substitute stationary
spherically-symmetric Ansatz 
\begin{equation}
  \label{eq:15}
  \varphi_Q(\boldsymbol{x}, \, t) = \chi_Q(r) \, \mathrm{e}^{i\omega
    t}\;, \qquad r = |\boldsymbol{x}|,
\end{equation}
into the classical field equations and numerically solve the remaining
ordinary differential equation for the real function $\chi_Q(r)$ with regularity
conditions   ${\partial_r   \chi_Q(0) =   \chi_Q(\infty) = 0}$, see
Appendix~\ref{sec:appendix-solitons} for details. This gives a family
of localized solutions parametrized with the frequency
$\omega$, see two of them in
Fig.~\ref{fig:Qw}a. 
In Fig.~\ref{fig:Qw}b we show the charges $Q = Q(\omega)$ of all 
solutions, Eq.~(\ref{eq:1}). Quite surprisingly, no solution exists at 
$Q<Q_c\approx 266/g^2$. At the same time, two solutions, the
``$Q$-ball'' and ``$Q$-cloud'', are present at each $Q>Q_c$; the
respective subfamilies are marked by solid and thick-dashed 
lines in Fig.~\ref{fig:Qw}b. 

\begin{figure}[htb]
  
  \includegraphics{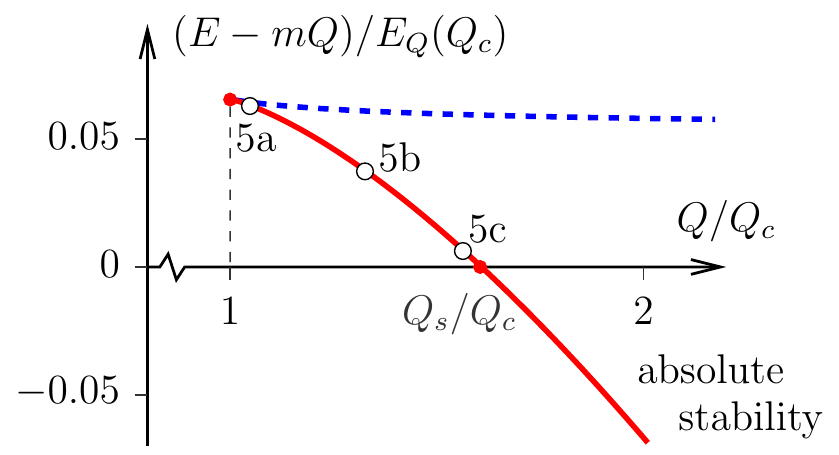} \hspace{1cm}
  \includegraphics{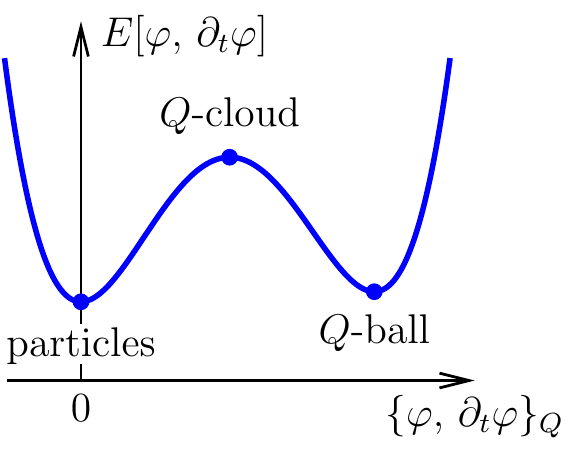}

  \hspace{4cm}(a) \hspace{7.3cm}(b)
  
  \caption{(a) Binding energies $E-m Q$ of $Q$-balls and $Q$-clouds as
    functions of their charges~$Q$ (solid and dashed lines,
    respectively). Empty circles show parameters of the tunneling
    solutions in Figs.~\ref{fig:solutions}a-c.
    (b) Schematic form of the energy functional $E[\varphi,\,
      \partial_t {\varphi}]$ in the subspace of configurations
    $\{\varphi(\boldsymbol{x}),\, \partial_t{\varphi}(\boldsymbol{x})\}_Q$
    with fixed charge $Q$. 
    \label{fig:EQ2}}
\end{figure}

To explain these features, we recall the main property of
nontopological solitons proven in~\cite{Coleman:1985ki}: 
stationary solutions of the form (\ref{eq:15}) are, in fact, extrema
of energy 
\begin{equation}
  \label{eq:21}
  E = \int d^3 \boldsymbol{x} \left[ \partial_t \varphi \partial_t
    \bar{\varphi} + 
    \nabla_{\boldsymbol{x}} \varphi \nabla_{\boldsymbol{x}} \bar{\varphi} +
    V(\varphi \bar{\varphi}) \right]
\end{equation}
in the subsector of field configurations $\{\varphi(\boldsymbol{x}),\,
\partial_t \varphi(\boldsymbol{x})\}_Q$ with fixed charge
$Q$. Computing the total masses (\ref{eq:21}) of all
solutions, we find that the $Q$-clouds are always heavier
than the $Q$-balls with the same charge, see
Fig.~\ref{fig:EQ2}a. Besides, the   $Q$-clouds are classically
unstable: they do not satisfy the rigorous Vakhitov-Kolokolov 
criterion $dQ/d\omega \leq 0$~\cite{Q-criterion,   Zakharov12,
  Panin:2016ooo} which is necessary for stability, cf.\ Fig.~\ref{fig:Qw}b. 
This suggests the structure of the energy functional in the subsector
of configurations with fixed $Q$ 
shown in Fig.~\ref{fig:EQ2}b. The $Q\mbox{-ball}$ and a configuration
describing $Q$ free particles at rest are the two local minima of $E$
at fixed  $Q$ separated by an energy barrier. The $Q$-cloud, to the 
contrary, is an unstable stationary solution ``sitting'' precisely on
the barrier top~\cite{Nugaev:2015rna}; the analogous solutions
in scalar theories with false vacuum decay and in gauge theories are
called critical bubble~\cite{Kobzarev:1974cp,   
  Coleman:1977py,    Coleman:1978ae} and
sphaleron~\cite{Klinkhamer:1984di}, respectively.

In Appendix \ref{sec:appendix-solitons} we confirm the picture in
Fig.~\ref{fig:EQ2}b 
by demonstrating that all $Q$-balls in our model are
classically stable, while the $Q$-clouds have precisely
one unstable mode.

In quantum theory the global minimum of energy in
Fig.~\ref{fig:EQ2}b corresponds to true ground state at a given $Q$,
while the other, local minimum represents metastable state decaying via
tunneling through the potential barrier. It is clear from
Fig.~\ref{fig:EQ2}a that the $Q$-balls are metastable at $Q_c <
Q<Q_s$ when their masses exceed the energies $mQ$ of free particles
with the same charge. Below we compute the lifetimes
$\Gamma_Q^{-1}$ of these objects.  

Direct inspectation of literature~\cite{Friedberg:1976me,
  Alford:1987vs, Lee:1991ax, Tsumagari:2008bv, Gulamov:2013ema, 
  Nugaev:2013poa, Nugaev:2015rna} shows that the
above picture is typical in weakly interacting theories with bounded
energies possessing $Q\mbox{-ball}$ solutions. Namely, nontopological solitons in
these models can be sorted into classically stable $Q$-balls~---
local minima of energy at fixed charge~--- and  $Q$-clouds
``sitting'' on tops of the potential barriers  between the 
$Q$-balls and free particles in the vacuum. The branches of 
$Q$-balls and $Q$-clouds meet at the critical point $Q=Q_c$, $E =
E_Q(Q_c)$ corresponding to the smallest soliton at the verge of
classical stability. By construction, the $Q$-cloud masses are higher than the
minimal values of energy $E_Q$ and $mQ$. Thus, the critical mass also   
satisfies $E_{Q}(Q_c) > mQ$ implying that there exists a window of
charges $Q_c < Q < Q_s$ where the $Q$-balls are metastable,  
$E_Q  > mQ$. 
  
\subsection{Euclidean solutions for the decay probability}
\label{sec:eucl-solut-decay-1}
Let us preview our semiclassical recipe to calculate $\Gamma_Q$
postponing its derivation till Sec.~\ref{sec:deriv-semicl-meth}. For
illustrative purposes we will compare this recipe to the powerful
Euclidean technique~\cite{Coleman:1977py,   Callan:1977pt,
  Coleman:1978ae} developed for the decay of metastable (false) vacuum in
models of real scalar field(s) $\phi$ without conserved
charges. The latter process is described by  the ``bounce''~--- real Euclidean
solution $\phi_{cl}(\boldsymbol{x}^2 +   \tau^2)$ interpolating
between the false vacuum at $\tau \to -\infty$ and the catchment  area
of the true vacuum at $\tau=0$. After continuation to 
Minkowski time $t  \equiv -i\tau > 0$ this solution describes
classical evolution of an expanding true vacuum bubble in the final
state.  The rate of false vacuum decay is given
by the expression  similar to Eq.~(\ref{eq:8}), where the leading
suppression exponent is Euclidean action of the bounce
$S_E[\phi_{cl}]$ and the prefactor includes fluctuation determinant
around this solution. 

We expect to find similar, though properly modified, procedure for
computing the rate of $Q$-ball decay. In
Sec.~\ref{sec:deriv-semicl-meth} we indeed  introduce the
semiclassical solution $\varphi_{cl}(\boldsymbol{x},\,  
\tau)$, $\bar{\varphi}_{cl}(\boldsymbol{x},\, \tau)$ satisfying 
Euclidean field equations in the model (\ref{eq:13}),
\begin{equation}
  \label{eq:6}
  (\partial_\tau^2 + \nabla_{\boldsymbol{x}}^2) \varphi_{cl}
  = V' \varphi_{cl}\;, \qquad 
  \qquad
  (\partial_\tau^2  + \nabla_{\boldsymbol{x}}^2)
  \bar{\varphi}_{cl} = V' \bar{\varphi}_{cl}\;,
\end{equation}
where $V'$ is a derivative of $V(\bar{\varphi} \varphi)$ with respect
to its argument. Our solution is spherically-symmetric, i.e.\ depends
on $r\equiv |\boldsymbol{x}|$ and $\tau$. It is natural to expect that
it coincides with the $Q$-ball~(\ref{eq:15}) at the start of the
process, 
\begin{equation}
  \label{eq:26}
  \varphi_{cl} \to \mathrm{e}^{\omega \tau
    - \eta_0/2}\chi_Q(r) \;, \qquad 
  \bar{\varphi}_{cl} \to \mathrm{e}^{-\omega \tau
    + \eta_0/2}\chi_Q(r)  \qquad \mbox{as}\qquad \tau \to
    -\infty \;,
\end{equation}
where we introduced the time shift $\eta_0$ that cannot be excluded in
general. In Sec.~\ref{sec:deriv-semicl-meth} we obtain the boundary
condition
\begin{equation}
  \label{eq:44}
  \varphi_{cl} = \mathrm{e}^{-\omega
    \beta-\eta_0}\,\bar{\varphi}_{cl}\;,\qquad
  \partial_\tau\varphi_{cl} = - \mathrm{e}^{-\omega
    \beta-\eta_0}\partial_\tau\bar{\varphi}_{cl} \qquad \mbox{at}
  \qquad \tau=-\frac{\beta}{2} \to -\infty\;,
\end{equation}
which is  consistent with the asymptotics (\ref{eq:26}). Soon we
will verify numerically that the solutions of
Eqs.~\eqref{eq:6}, (\ref{eq:44}) automatically satisfy
Eq.~(\ref{eq:26}).

Surprisingly,  $\varphi_{cl}$ and
$\bar{\varphi}_{cl}$ in Eq.~\eqref{eq:26} are not complex  conjugate
to each other. This is not a problem, however, for the complex
semiclassical method adopted in
Sec.~\ref{sec:deriv-semicl-meth}. Instead, our derivation suggests
that $\varphi_{cl}$ and $\bar{\varphi}_{cl}$ should be considered as
independent {\it real} functions of $\boldsymbol{x}$ and $\tau$. Then
the value of Euclidean action 
\begin{equation}
\label{eq:S_E}
S_E \equiv -iS =  \int_{-\beta/2}^{\beta/2} d\tau d^3
\boldsymbol{x} 
\left[\partial_\tau\varphi \partial_\tau\bar{\varphi}  +
  \nabla_{\boldsymbol{x}} \varphi \nabla_{\boldsymbol{x}}
  \bar{\varphi}  + V(\varphi \bar{\varphi}) \right] 
\end{equation}
is real, where $\beta$ will be sent to infinity in the end of the
calculation.
\begin{sloppy}

Like the bounce, our semiclassical solution arrives to the catchment
area of the true vacuum at $\tau=0$. To quantify this criterion, we
use the second boundary condition derived in 
Sec.~\ref{sec:deriv-semicl-meth},
\begin{equation}
  \label{eq:43}
  \varphi_{cl} = \bar{\varphi}_{cl}\;,\qquad \partial_\tau\varphi_{cl}
  = -\partial_\tau\bar{\varphi}_{cl} \qquad  \mbox{at} \qquad \tau=0\;.
\end{equation}
This means that  the solution is symmetric with
respect to time reflections, ${\varphi_{cl}(\boldsymbol{x},\,   \tau)
  = \bar{\varphi}_{cl}(\boldsymbol{x},\, -\tau)}$. As a consequence,
after continuation to Minkowski time ${t     = -i\tau > 0}$ the
functions $\varphi_{cl}$ and $\bar{\varphi}_{cl}$ become complex
conjugate to each other, i.e.\ represent ordinary 
classical evolution after the decay. We expect to obtain free
waves/particles classically evolving in the vacuum as~${t\to
  +\infty}$.

\end{sloppy}
Solving Eqs.~\eqref{eq:6} with boundary conditions~\eqref{eq:44},
\eqref{eq:43} at large $\beta$, one finds a family of real  solutions
$\varphi_{cl}$, $\bar{\varphi}_{cl}$ parametrized with $\omega\beta +
\eta_0$ in Eq.~(\ref{eq:44}). Apparently, different values of this
combination correspond to solutions approaching different $Q$-balls at
large negative $\tau$. In what follows we express $\eta_{0} = \eta_{0}(Q)$
using Eq.~\eqref{eq:1} and characterize the solutions with 
charge~$Q$. Sending $\beta \to +\infty$, we obtain $\varphi_{cl}$
and $\bar{\varphi}_{cl}$
interpolating between the $Q$-balls~\eqref{eq:26} at $\tau\to -\infty$
and the sector of true vacuum at $\tau=0$. After that the solutions are continued
to $\tau > 0$ using the time reflection symmetry. 

Given $\varphi_{cl}$, $\bar{\varphi}_{cl}$, $\eta_0$,
one computes the suppression exponent  
\begin{equation}
  \label{eq:25}
  F_Q = S_E[\varphi_{cl},\, \bar{\varphi}_{cl}] - S_E[\varphi_{Q},\,
    \bar{\varphi}_{Q}] + \eta_0 Q
\end{equation}
of the decay rate~(\ref{eq:8}). Note that the difference of Euclidean
actions (\ref{eq:S_E}) on $\varphi_{cl}$, $\bar{\varphi}_{cl}$ and 
$Q$-ball (\ref{eq:15}) in this expression remains finite as  $\beta
\to +\infty$ because our solutions approach $\varphi_{Q}$, 
$\bar{\varphi}_Q$ at large $|\tau|$.  The third term in
Eq.~(\ref{eq:25}) is specific to  models with conserved  charge
$Q$. Notably, it involves the parameter $\eta_0$ from
Eq.~(\ref{eq:44}) at the place typically occupied by the chemical potential
of $Q$ in statistical physics. In
Sec.~\ref{sec:saddle-point-appr} we derive the Legendre transform
formula ${dF_Q /dQ = \eta_0}$ supporting this intuition.

Our expression for the prefactor in Eq.~(\ref{eq:8}) is way more
complicated. It uses small perturbations in the background of the
semiclassical solution, 
$$
\varphi = \varphi_{cl}  + \mathrm{e}^{\omega \tau} \delta \varphi\;,
\qquad
\bar{\varphi} = \bar{\varphi}_{cl}  + \mathrm{e}^{-\omega \tau} \delta
\bar{\varphi}\;,
$$
where $\mathrm{e}^{\pm \omega\tau}$ compensate for
exponential behavior of $\varphi_{cl}$ and $\bar{\varphi}_{cl}$ as
$\tau\to \pm \infty$. The effect of perturbations on the Euclidean
action is characterized by its second variation, i.e.\ the operator
\begin{equation}
  \label{eq:42}
  \hat{\cal D}_{cl} = \frac{\delta^2 S_E}{(\delta \varphi,\,
    \delta \bar{\varphi})^2} \Bigg|_{cl}
  = \begin{pmatrix} V''_{cl} \bar{\chi}_{cl}^2 & \hat{\cal D}^{-}
    \\ \hat{\cal D}^{+} & V''_{cl} 
    \chi^2_{cl}\end{pmatrix} \;, \;\; \hat{\cal D}^{\pm} \equiv
    -(\partial_\tau\pm \omega)^2 - \nabla_{\boldsymbol{x}}^2 + V'_{cl} + V''_{cl}
    \chi_{cl}\bar{\chi}_{cl}\;.
\end{equation}
Here we introduced the rescaled fields $\chi_{cl} \equiv
\mathrm{e}^{-\omega\tau}\, \varphi_{cl}$ and $\bar{\chi}_{cl} \equiv
\mathrm{e}^{\omega\tau} \, \bar{\varphi}_{cl}$ remaining finite as
$\tau \to \pm \infty$ and denoted $V_{cl} \equiv V(\varphi_{cl} 
\bar{\varphi}_{cl})$; the primes are derivatives of this function
with respect to its argument. Note that $\hat{\cal D}_{cl}$ is Hermitean:
it acts on the perturbations ${(\delta\varphi,\,
  \delta\bar{\varphi})^{T}}$ vanishing at $\tau \to \pm
\infty$. The prefactor in Eq.~(\ref{eq:8}) involves determinant of this
operator, 
\begin{equation}
  \label{eq:57}
   A_Q  = \cosh^2\eta_0 \,\left(\frac{B_0}{2\pi}\right)^{1/2}
  \left[\frac{-\mathrm{det}'  \,
      \hat{\cal D}_{cl}}{\phantom{-}\mathrm{det}' \, \hat{\cal
        D}_Q}\right]^{-1/2}\;, 
\end{equation}
where $\det'$ means that all zero eigenvalues are excluded from the
determinant, $\hat{\cal D}_Q$ is the same as $\hat{\cal D}_{cl}$ but
in the $Q$-ball background i.e.\ with $\chi_{cl},\, \bar{\chi}_{cl}
\to \chi_Q(r)$ in Eq.~(\ref{eq:42}), the factors $\cosh^2 \eta_0$ and 
$B_0 = 2 \int d\tau d^3 \boldsymbol{x}\; (\partial_\tau \chi_{cl})^2$
were introduced while deleting zero modes from the 
determinants. Like in the case of bounce, we expect that
$\mathrm{det}'\, \hat{\cal D}_{cl}$ has opposite sign to the same
determinant $\mathrm{det}'\,\hat{\cal D}_Q$ computed in the background
of the metastable solution.

Expression~(\ref{eq:57}) for the prefactor seems too involved for 
practical calculations (see, however,~\cite{Callan:1977pt,
  Baacke:2003uw, Dunne:2005rt,Andreassen:2016cvx}). But it is
straightforward to estimate $A_Q$ by extracting its dependence on the only
small parameter in the model~--- the semiclassical parameter $g^2
\propto Q_c^{-1}$. The determinants in Eq.~(\ref{eq:57}) do
not depend on $g^2$, while $B_0 \propto g^{-2} \sim Q_c$. One
obtains $A_Q = mQ_c^{1/2}$, where the mass $m$ is restored on
dimensional grounds.

\subsection{Applying the method}
\label{sec:applying-formula}
\begin{figure}[htb]
  \centerline{
    \includegraphics[trim=0 3 0 0, scale=.94] {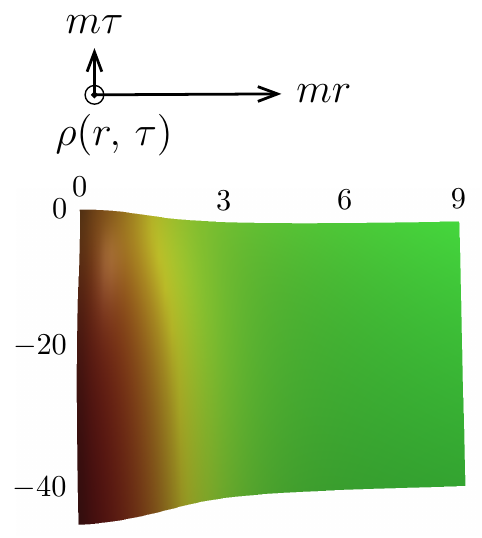}
    \includegraphics[scale=.94] {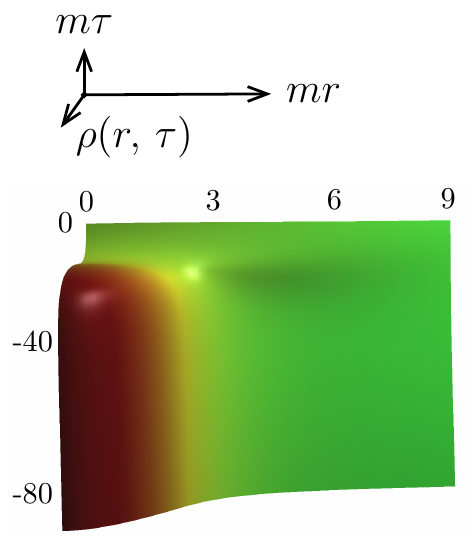}\hspace{-2mm}
    \includegraphics[scale=.94] {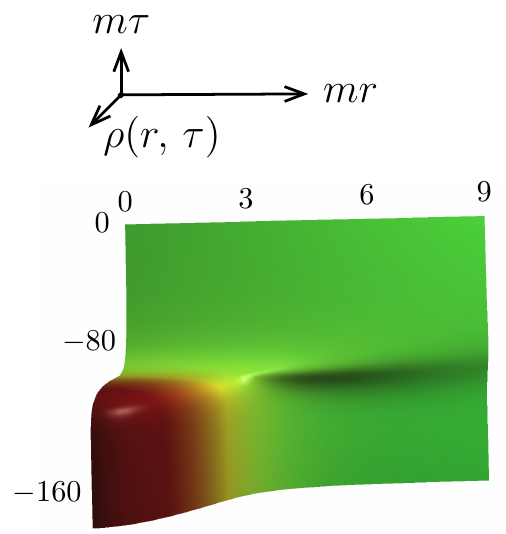}\hspace{1mm}
    \includegraphics[scale=.94] {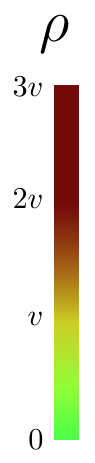}}


  \hspace{2.3cm}(a) \hspace{4.2cm}(b) \hspace{4cm} (c)
    \caption{Semiclassical solutions $\rho(r,\, \tau)\equiv (\varphi_{cl}
      \bar{\varphi}_{cl})^{1/2}$ describing  decay of $Q$-balls with $Q/Q_c
      \approx$ $1.05$~(a), $1.33$~(b), and
      $1.56$~(c). The parameters of these solutions are marked by 
      open circles in Fig.~\ref{fig:EQ2}a.\label{fig:solutions}}
\end{figure}

\begin{figure}[htb]

  \hspace{3.7cm} (a) \hspace{7cm} (b)
    
  \centerline{\includegraphics[scale=.95]{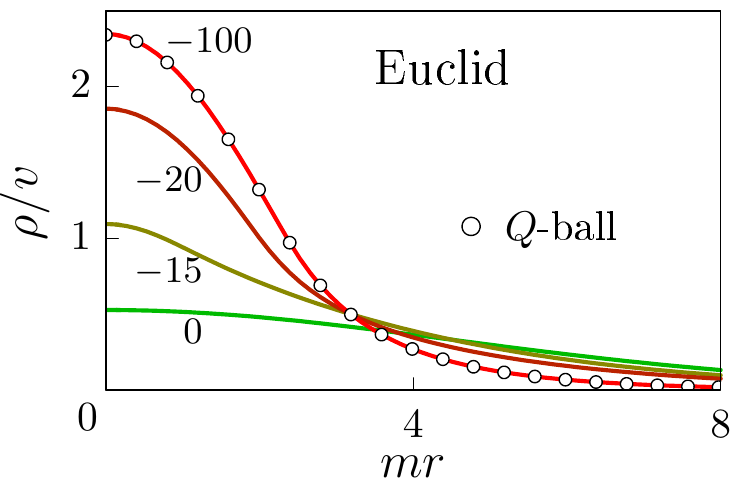} \hspace{2mm}
    \includegraphics[scale=.95]{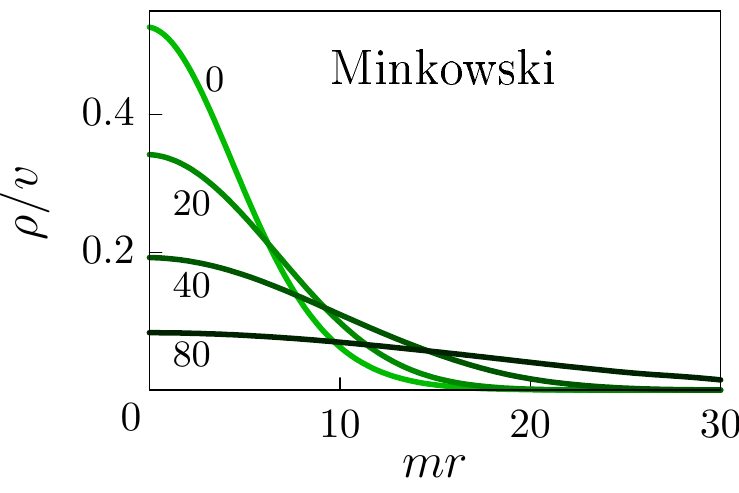}}
  \caption{(a)~Sections $\tau = \mbox{const}$ of the semiclassical
    solution $\rho(r,\, \tau) \equiv (\varphi_{cl}\bar{\varphi}_{cl})^{1/2}$ in
    Fig.~\ref{fig:solutions}b (lines). The values of $m\tau$ are
    written near the graphs. Empty circles represent the $Q$-ball
    profile $\chi_Q(r)$ at the same charge. (b)~The same semiclassical 
    solution continued to Minkowski time $t = -i\tau$. Solid lines show
    $\rho(r,\, t) \equiv     |\varphi_{cl}|$ at fixed $mt$
    (numbers near the 
    graphs).\label{fig:sections}}  
\end{figure}

Now, let us illustrate the semiclassical method in the model (\ref{eq:2}) leaving
technical details to Sec.~\ref{sec:numer-impl}. We numerically  
find the family of semiclassical solutions satisfying
Eqs.~(\ref{eq:6}), (\ref{eq:44}) and  (\ref{eq:43}),
then compute their charges by Eq.~(\ref{eq:1}).
Figure~\ref{fig:solutions} 
demonstrates profiles $\rho(r,\, \tau)\equiv (\varphi_{cl}
\bar{\varphi}_{cl})^{1/2}$ of three solutions from this family; the
sections $\tau = \mbox{const}$ of the solution in
Fig.~\ref{fig:solutions}b are shown in Fig.~\ref{fig:sections}a.
Notably, all our solutions start from the stationary $Q$-ball 
$\chi_Q(r)$ (points in Fig.~\ref{fig:sections}a) at $\tau \to -\infty$
and become lower and wider at $\tau =0$ where $\varphi_{cl}
\bar{\varphi}_{cl} < v^2$. To confirm that they indeed arrive to the
vacuum sector, we
continue $\varphi_{cl}$ and $\bar{\varphi}_{cl}$ to 
real time by numerically solving the field 
equations from $t = \tau = 0$ to large $t$, see
Fig.~\ref{fig:sections}b. As expected, Minkowski evolution of 
our solutions describes spreading wave packets
corresponding to free particles in the vacuum. 

\begin{figure}[htb]

  \vspace{7mm}
  \hspace{3.5cm} (a) \hspace{7.3cm}(b)

  \vspace{-7mm}
  \centerline{\includegraphics{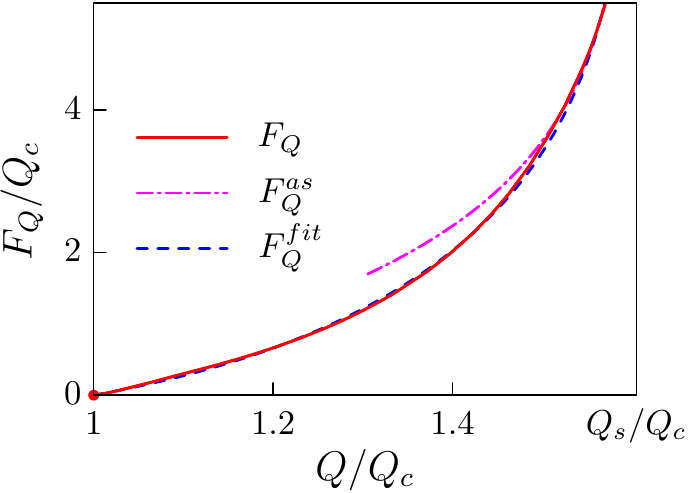} \hspace{3mm}
    \includegraphics{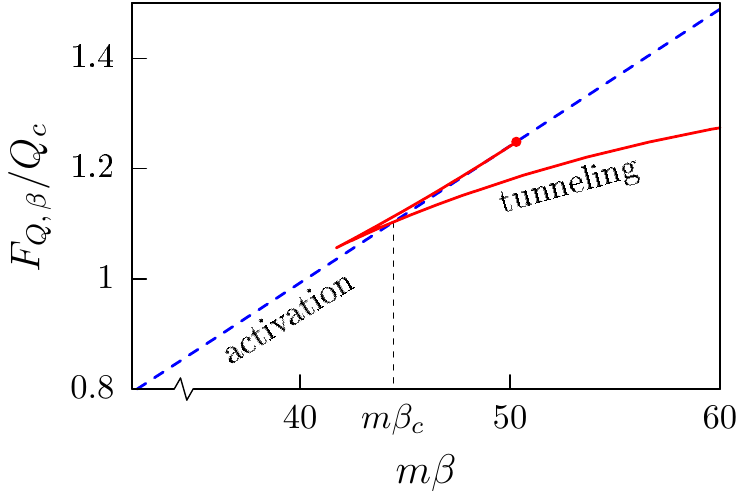}}
  \caption{(a) Suppression exponent $F_Q$ for the rate of $Q$-ball
    decay, its $Q\to Q_s$ asymptotic $F_Q^{as}$ and fitting
    function~$F_Q^{fit}$, Eqs.~(\ref{eq:28}) and (\ref{eq:32}).
    (b) The exponent $F_{Q,\, \beta}$ of thermally induced decay as a
    function of the inverse temperature $\beta \equiv T^{-1}$ at  
    $Q\approx 1.33\, Q_c$. Solid and diagonal-dashed lines are computed 
    using the quasiperiodic Euclidean solutions and $Q$-cloud,
    respectively. The true suppression exponent is given by the least
    suppressed contribution at each $\beta$, i.e.\  the  lowest line on
    the plot.\label{fig:Fbeta}} 
\end{figure}

Given the solutions, we numerically compute the suppression
exponent~(\ref{eq:25}) (solid lines in
Figs.~\ref{fig:result}b and  \ref{fig:Fbeta}a). We see that $F_Q$
equals zero and infinity at charges $Q_c$ and $Q_s$
corresponding to classically unstable and absolutely stable 
$Q$-balls, respectively. 

To find practical fitting formula for $F_Q$, let us study its  behavior
at $Q\to Q_s$ when the total energy deficit $Q
\epsilon_p \equiv E_Q - mQ$ 
of the decay is small. This corresponds to small momenta $p \sim
(2m\epsilon_p)^{1/2} \propto (1 -  Q/Q_s)^{1/2}$ of the
final-state particles/waves 
and, as a consequence, slow 
evolution of the solution near $\tau=0$: $\bar{\varphi}_{cl}\varphi_{cl} \propto
\mathrm{e}^{\tau\epsilon_p}$. Thus, large time delay $\eta_0 
\propto \epsilon_p^{-1} \propto (1 - Q/Q_s)^{-1}$ is needed for
$\varphi_{cl} \bar{\varphi}_{cl}$ to approach the
$Q$-ball profile. Integrating the Legendre formula $dF_Q/dQ = \eta_0$, we 
obtain the asymptotic of the suppression exponent
\begin{equation}
  \label{eq:28}
  F_Q \to  F_Q^{as} \equiv d_1 + d_2 \log \left(1 - Q/Q_s  \right)
  \qquad \mbox{as} \qquad Q\to Q_s\;.
\end{equation}
Figure~\ref{fig:Fbeta}a demonstrates that Eq.~(\ref{eq:28}) (dash-dotted
line) is indeed close to $F_Q$ at ${Q \to Q_s}$ for properly
chosen $d_i$. Slightly generalizing Eq.~(\ref{eq:28}), we obtain 
the fitting formula in the entire metastability window $Q_c < Q < Q_s$, 
\begin{equation}
  \label{eq:32}
  F_Q \approx F_Q^{fit} = (Q-Q_c) \left[ c_1 + c_2 \log \left(1 - Q/Q_s  \right)
    \right]\;.
\end{equation}
We find that Eq.~(\ref{eq:32}) with $c_1=-0.28$ and $c_2=-2.6$
describes our numerical results 
with $5\%$ relative precision, see the dashed line in
Fig.~\ref{fig:Fbeta}a.  

\subsection{Decay at finite temperature}
\label{sec:decay-at-finite-1}
Consider metastable $Q$-ball immersed into neutral plasma at 
temperature~$T$. Thermal fluctuations should decrease
the lifetime $\Gamma_{Q,\, \beta}^{-1}$ of this object kicking it 
over the potential barrier in Fig.~\ref{fig:EQ2}b. Fortunately, our
semiclassical method is easily generalized to the case of finite temperature $T
\equiv \beta^{-1}$: one just considers the above semiclassical
solutions in a finite time interval $-\beta/2 < \tau < \beta/2 $ and imposes
Eq.~(\ref{eq:44}) at $\tau = -\beta/2$. This makes the functions
$\varphi_{cl}$ and $\bar{\varphi}_{cl}$ quasiperiodic: their values at
$\tau = \pm \beta/2$ differ  by  multiplicative factors
$\mathrm{e}^{\pm (\omega\beta +   \eta_0)}$. The suppression exponent
is still given by Eq.~(\ref{eq:25}), where $S_E$ is the Euclidean
action~\eqref{eq:S_E} 
in the interval $|\tau| < \beta/2$. Generalization of the
prefactor formula (\ref{eq:57}) is slightly less trivial, see comments
in Sec.~\ref{sec:decay-at-finite}. 

We numerically obtained the quasiperiodic solutions in the
model (\ref{eq:2}). Their suppression exponent $F_{Q,\,   \beta}$ is
represented by the solid (``tunneling'') line in Fig.~\ref{fig:Fbeta}b. Notably, 
the solutions of this kind do not exist at small $\beta$ and therefore
fail to describe thermal transitions in that region. To consider
high-temperature processes, we include a trivial semiclassical solution~--- 
the $Q$-cloud~--- which has the same form~(\ref{eq:15}) as the
$Q$-ball of the same charge,
but with different frequency $\tilde{\omega}$ and profile
$\tilde{\chi}_Q(r)$.  This solution satisfies Eqs.~(\ref{eq:6}), (\ref{eq:44}),
(\ref{eq:43}) at arbitrary $\beta$. Instead of interpolating between the sectors of
$Q$-ball and the true vacuum, it simply sits on top of the barrier between
them, see Fig.~\ref{fig:EQ2}b. Computing the exponent~(\ref{eq:25})
for the $Q$-cloud, one obtains linear function $F_{Q,\, \beta} = \beta
(\tilde{E}_Q - E_Q)$ shown by the diagonal dashed line in  Fig.~\ref{fig:Fbeta}b,
where $\tilde{E}_Q$ is the mass of this object.

The picture in Fig.~\ref{fig:Fbeta}b is typical~\cite{Kuznetsov:1997sf} for thermal
transitions, with two semiclassical contributions to the rate
representing direct tunneling through the potential barrier in  
Fig.~\ref{fig:EQ2}b and jumps onto its top (activation) induced by
thermal fluctuations. The overall suppression exponent 
corresponds 
to the least suppressed contribution at each $\beta$,
given by the ``tunneling'' and ``activation'' lines in Fig.~\ref{fig:Fbeta}b
at temperatures lower and higher $\beta_c^{-1}$, respectively. 

As a further generalization of the method, one may consider decay of
$Q$-ball in charged plasma at temperature $T$ and chemical
potential $\mu$. This process, however, is different from the physical
viewpoint~\cite{Laine:1998rg}. The $Q$-ball exchanges charge with the
medium and reaches equilibrium at $Q = Q_\mu$. The respective
initial state is described by the grand canonical ensemble giving 
$$
\Gamma_{\mu,\, \beta} \propto \int dQ \;\mathrm{e}^{\mu
    (Q - Q_\mu) -F_{Q,\, \beta}}
$$
for the decay rate, where the prefactors are ignored. This integral is
saturated either by one of the integration limits or 
in the vicinity of the saddle point $\mu = \eta_0$; one has to select
the least suppressed contribution at every 
$\mu$ and $\beta$. We leave this interesting calculation to future
studies.

\section{Derivation of the semiclassical method}
\label{sec:deriv-semicl-meth}
We are going to derive the semiclassical expression for the $Q$-ball
decay rate in two steps. First, we write this rate in the form of 
path integral.  Second, we evaluate the integral in the
saddle-point approximation. We will see that this technique is 
trustworthy at~$Q\gg 1$.

\subsection{Quantum states for $Q$-balls}
\label{sec:quantum-q-ball}
The first nontrivial step of our program is to {\it define} the states
of quantum $Q$-balls. Note that identification of these objects with
the classical solutions (\ref{eq:15}) is not satisfactory in quantum
theory where the $Q$-ball state $|Q\rangle$ should be
explicitly specified in order to compute its decay amplitude. 

Our definition of $Q$-balls essentially relies on the presence of 
a conserved charge in the theory. In the simplest model
(\ref{eq:13}) the charge $Q$ is associated with the phase rotation symmetry
$\varphi \to \mathrm{e}^{-i\alpha} \varphi$. Then in quantum case
the operator $\hat{Q}$ generates symmetry transformations, i.e.\ acts
as 
\begin{equation}
  \label{eq:4}
  \mathrm{e}^{i\alpha \hat{Q}} | \varphi,\, \bar{\varphi}\rangle  =
  |\mathrm{e}^{-i\alpha} \varphi ,\, \mathrm{e}^{i\alpha}
  \bar{\varphi} \rangle
\end{equation}
on the eigenstates $|\varphi,\, \bar{\varphi}\rangle$ of field
operators $\hat{\varphi}$, $\hat{\bar{\varphi}}$ with eigenvalues  
$\varphi(\boldsymbol{x})$ and $\bar{\varphi}(\boldsymbol{x})$. Since
$\mathrm{e}^{2\pi i \hat{Q}}= 1$ corresponds to full phase rotation, the
charge has integer eigenvalues. Then the operator 
\begin{equation} 
  \label{eq:5}
  \hat{P}_Q =  \int\limits_0^{2\pi i} \frac{d\eta}{2\pi i} \;\;
  \mathrm{e}^{\eta (\hat{Q}-Q)}\;, \qquad \qquad \hat{P}_Q^2 = \hat{P}_Q\;,
\end{equation}
projects onto the subspace of states with given $Q$; note that
integration in Eq.~(\ref{eq:5}) is performed along imaginary
$\eta$. 

Using the projector (\ref{eq:5}) one can study the subsector of states
with fixed~$Q$. It is natural to define quantum $Q$-ball
as a state $|Q\rangle$ of minimal energy within this
subsector~\cite{Coleman:1985ki}. We obtain
the limiting formula 
\begin{equation}
  \label{eq:7}
  \mathrm{e}^{-\beta\hat{H}/2 } \hat{P}_Q| i\rangle  \to \mathrm{e}^{-\beta E_Q /2
    }   | Q\rangle  \langle Q | i \rangle
  \qquad \mbox{as} \qquad \beta\to +\infty\;, 
\end{equation}
where $\hat{P}_Q$ projects an arbitrary state $|i\rangle$ onto the
charge-$Q$ subsector, while the operator
$\mathrm{e}^{-\beta\hat{H}/2}$ at large $\beta$ suppresses all excited
states within this subsector leaving only the minimal-energy 
eigenstate $|Q\rangle$ of the Hamiltonian $\hat{H}$. For   
simplicity we use normalization\footnote{Assuming that a finite
  spatial box with some 
  boundary conditions is introduced.}~${\langle Q|Q\rangle =1}$.  

In fact, we are interested in small $Q$-balls representing local
(false) minima of energy at fixed $Q$, see Fig.~\ref{fig:EQ2}b. One
can expect that 
Eq.~(\ref{eq:7}) is still applicable in this case if the initial state
$|i\rangle$ belongs to the catchment area of the Q-ball and Euclidean time
interval $\beta$ is not exponentially large.

\subsection{Decay probability in the form of path integral}
\label{sec:path-integral-decay}
The total probability of Q-ball decay in time  $t_0$ is obtained by
time-evolving the state $|Q\rangle$ and projecting it onto the basis
of Fock states $|f\rangle$ above the vacuum,
\begin{equation}
  \label{eq:3}
  {\cal P} = \sum_f  \left|
  \langle f  | \mathrm{e}^{-i\hat{H} t_0}
  | Q\rangle\right|^2  = \mathrm{e}^{\beta E_Q} \sum_{i,\, f} \;\left|
  \langle f|  \mathrm{e}^{-i\hat{H}(t_0- i\beta/2)  } \hat{P}_Q | 
  i\rangle\right|^2\;,
\end{equation}
where Eq.~(\ref{eq:7}) was used to obtain  the second equality and the
limits $\beta\to +\infty$, $t_0 \to +\infty$ are assumed from now
on. Importantly, all 
initial and final states $|i \rangle$ and $|f\rangle$ in
Eq.~(\ref{eq:3}) belong to the catchment areas of the $Q$-ball and 
true vacuum, respectively. In particular, due to unitarity of quantum
evolution one would obtain unit probability,
\begin{equation}
  \label{eq:9}
  \mathrm{e}^{\beta E_Q} \sum_{i,\, \Psi} \;\left|
  \langle \Psi|  \mathrm{e}^{-i\hat{H}(t_0 - i\beta/2)  } \hat{P}_Q | 
  i\rangle\right|^2 \to 1 \qquad \mbox{as} \qquad \beta\to +\infty\;,
\end{equation}
if summation over all Hilbert states $|\Psi\rangle$ was performed instead of
$|f\rangle$. 

\begin{figure}[htb]
  \centerline{
    \includegraphics{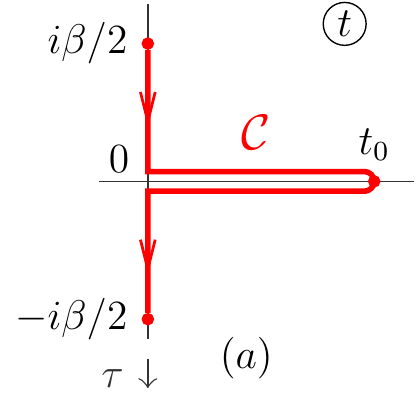}
    \hspace{3cm}
    \includegraphics{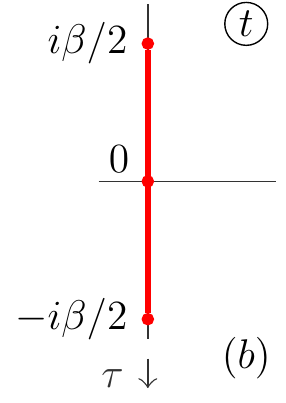}\hspace{5mm}}
  \caption{(a) Time contour for the path
    integral~\eqref{eq:11}. (b) Equivalent contour.\label{fig:contour}}
\end{figure}

To write path integral for the probability (\ref{eq:3}), we use the
basis of configuration eigenstates $|i\rangle  \equiv  |\varphi_i,\,
\bar{\varphi}_i\rangle$, $|f\rangle \equiv |\varphi_f,\,
\bar{\varphi}_f\rangle$ and Eqs.~\eqref{eq:4}, \eqref{eq:5} for
the projector\footnote{We also exploited the property $\sum_i
  \hat{P}_Q |i \rangle \langle i | \hat{P}_Q = \sum_i
  \hat{P}_Q|i\rangle \langle i |$ which is valid up to exponentially
  small corrections because the states $|i \rangle$ belong to the
  $Q$-ball sector, cf.~Eq.~(\ref{eq:5}).}
\begin{multline}
  \notag
        {\cal P} = \int\limits_0^{2\pi i} \frac{d\eta}{2\pi i} \,
        \mathrm{e}^{\beta E_Q - \eta Q}  \int {\cal D}\varphi_i \,
               {\cal D}\bar{\varphi}_i\,  {\cal D} \varphi_f \,{\cal
                 D}\bar{\varphi}_f \;
               \langle \varphi_i ,\, \bar{\varphi}_i |
        \mathrm{e}^{-i\hat{H}(-i\beta/2-t_0)  }| \varphi_f
        ,\, \bar{\varphi}_f \rangle
        \\\times
        \langle \varphi_f ,\, \bar{\varphi}_f |
        \mathrm{e}^{-i\hat{H}(t_0- i\beta/2)  }|\mathrm{e}^{-\eta} \varphi_i
        ,\, \mathrm{e}^{\eta}\bar{\varphi}_i \rangle\;.
\end{multline}
One can interpret the propagation operators in this expression as
describing time evolution from $t = i\beta/2$ to $t = t_0$ to $t = -i\beta/2$
along the complex time contour ${\cal C}$ in Fig.~\ref{fig:contour}a. 
Using path integral representation for the evolution operators,
one finally arrives at expression,
\begin{equation}
  \label{eq:11}
        {\cal P} = \int\limits_0^{2\pi i} \frac{d\eta}{2\pi i} \;
        \mathrm{e}^{\beta E_Q - \eta Q}  \int {\cal D}\varphi(x)\, 
               {\cal D}\bar{\varphi}(x)\;  \mathrm{e}^{iS[\varphi,\,
                   \bar{\varphi}]}\;,
\end{equation}
where the configurations $\varphi(\boldsymbol{x},\, t)$,
$\bar{\varphi}(\boldsymbol{x},\, t)$ live on the 
contour ${\cal C}$ and satisfy quasi-periodic conditions
\begin{equation}
  \label{eq:12}
  \varphi(\boldsymbol{x},\, t + i\beta)  = \mathrm{e}^{-\eta}\varphi(
  \boldsymbol{x}, \, t)\;, \qquad  \bar{\varphi}(
  \boldsymbol{x},\, t + i\beta) =
  \mathrm{e}^{\eta}\bar{\varphi}(\boldsymbol{x},\, t)
\end{equation}
at $t = -i\beta/2$. The classical action $S[\varphi,\,
  \bar{\varphi}]$ is computed along 
the same contour. Note that the functions $\varphi$ and
$\bar{\varphi}$ can be continued to the entire Euclidean axis using
Eq.~(\ref{eq:12}).

Recall that the Euclidean parts of the contour ${\cal C}$ in
Fig.~\ref{fig:contour}a are not related to Wick
rotation, they appeared due to the minimization
procedure~\eqref{eq:7}. Thus, representation
(\ref{eq:11}) does not   
rely on analytic properties of the $Q$-ball decay amplitude,
cf.~\cite{Coleman:1978ae}. Note also that that the ``final-state''
configurations $\varphi(\boldsymbol{x},\, t_0) \equiv
\varphi_f(\boldsymbol{x})$ in Eq.~(\ref{eq:11}) should describe free
particles in the vacuum; this makes our integral for the 
probability essentially different from that for the unity
(\ref{eq:9}). Likewise, $\varphi(\boldsymbol{x},\, \pm i\beta/2)$
should belong to the catchment area of the $Q$-ball.

\subsection{Saddle-point approximation}
\label{sec:saddle-point-appr}
In Sec.~\ref{sec:small-q-balls} we argued that the action $S$ and
charge $Q$ take large values in the semiclassical regime $g^2 \ll 
1$. In this case the integral~(\ref{eq:11}) of the fast-oscillating
exponent  $\mathrm{e}^{iS}$ can be evaluated in the saddle-point
approximation.

We introduce the saddle-point configuration
$\{\bar{\varphi}_{cl},\, \varphi_{cl},\, \eta_{cl}\}$ as an extremum of
the integrand in Eq.~\eqref{eq:11}. It will be convenient 
to consider  $\varphi_{cl}(\boldsymbol{x},\, \tau)$ and
$\bar{\varphi}_{cl}(\boldsymbol{x},\, \tau)$ as functions of Euclidean 
time $\tau \equiv it$ which takes real and imaginary values on the
contour in Fig.~\ref{fig:contour}a.  By construction, $\varphi_{cl}$
and $\bar{\varphi}_{cl}$ satisfy the field equations (\ref{eq:6}) and
the boundary conditions~\eqref{eq:12},
\begin{equation} 
  \label{eq:40}
  \varphi_{cl}(\boldsymbol{x},\, \tau + \beta)  =
  \mathrm{e}^{\eta_{cl}}\varphi_{cl}( 
  \boldsymbol{x}, \, \tau)\;, \qquad  \bar{\varphi}_{cl}(
  \boldsymbol{x},\, \tau + \beta) =
  \mathrm{e}^{-\eta_{cl}}\bar{\varphi}_{cl}(\boldsymbol{x},\, \tau)\;.
\end{equation}
Besides, extremization with respect to $\eta$ gives
equation 
\begin{equation}
  \label{eq:17}
  Q = i\, \frac{d S}{ d\eta} = \int d^3 \boldsymbol{x} \left[
    \bar{\varphi}_{cl} \partial_\tau \varphi_{cl} - \varphi_{cl}
    \partial_\tau \bar{\varphi}_{cl}\right] \;,
\end{equation}
where Eqs.~(\ref{eq:6}), (\ref{eq:40}) were used in the
differentiation. Note that Eq.~(\ref{eq:17}) coincides with 
the classical expression~(\ref{eq:1}) for the  charge. Importantly, 
$\varphi_{cl}$ and $\bar{\varphi}_{cl}$ are not necessarily complex
conjugate to each other: integrations in Eq.~(\ref{eq:11}) can be
continued to independent $\varphi$- and $\bar{\varphi}$-contours in
the functional space.

Solving the field equations \eqref{eq:6} with the quasiperiodicity conditions
(\ref{eq:40}), one obtains the saddle-point configuration $\{
\varphi_{cl},\, \bar{\varphi}_{cl}\}$ for every $\eta_{cl}$. The
latter parameter is then related to the conserved charge by
Eq.~(\ref{eq:17}). 
\begin{sloppy}

Now, we compute the integral (\ref{eq:11}) by noting that at $g^2 \ll
1$ ($S \gg 1$) its integrand is sharply peaked in small vicinity of the
saddle point, 
\begin{equation} 
  \label{eq:27}
  \varphi = \varphi_{cl} +  \mathrm{e}^{\omega \tau}\, \delta \varphi\;,
  \qquad
  \bar{\varphi} = \bar{\varphi}_{cl} +
  \mathrm{e}^{- \omega \tau}\, \delta \bar{\varphi}\;, \qquad
  \eta = \eta_{cl} + \delta \eta\;.
\end{equation}
Here the factors $\mathrm{e}^{\pm \omega \tau}$ in
front of field  
perturbations compensate for exponential  growth of  $\varphi_{cl}$ and
$\bar{\varphi}_{cl}$ as ${\tau \to \pm\infty}$,
cf.\ Eq.~\eqref{eq:26}. Substituting Eq.~(\ref{eq:27}) into
Eq.~(\ref{eq:11}) and taking the Gaussian integrals over small $\delta
\varphi(\boldsymbol{x},\, \tau)$, $\delta
\bar{\varphi}(\boldsymbol{x},\, \tau)$ and $\delta \eta$,  one obtains
the standard saddle-point formula,
\begin{equation}
  \label{eq:41}
        {\cal P} = N\cdot \mathrm{e}^{\beta E_Q  - \eta_{cl} Q +
          iS[\varphi_{cl},\,  \bar{\varphi}_{cl}]} \cdot  
        \frac{\mathrm{det}^{-1/2}\hat{\cal
            D}_{cl}}{\sqrt{dQ/d\eta_{cl}}} \;.
\end{equation}
Here $N$ is the unknown normalization factor from the functional
measure, the second multiplier is the  saddle-point value of
the integrand, while the determinant of $\hat{\cal D}_{cl}$ defined in  
Eq.~\eqref{eq:42} and the factor ${dQ/d\eta_{cl} =  i d^2 S/d\eta^2}$
account for fluctuations of  $\{\delta \varphi,\,  \delta \bar{\varphi}\}$
and $\delta \eta$, respectively. 

\end{sloppy}
Now, let us us simplify the above semiclassical recipe. First, the
double-bent Minkowskian part of the contour ${\cal C}$ is
redundant: analytic functions $\varphi_{cl}$ and $\bar{\varphi}_{cl}$
can be considered on the Euclidean time axis in
Fig.~\ref{fig:contour}b. There 
remains, however, an important selection rule: when continued to 
real time, these functions should describe the final state with free
particles at ${t = t_0\to +\infty}$.

Second, the saddle-point equations have
two discrete symmetries  at $\tau \in \mathbb{R}$: simultaneous
complex conjugation  of all fields $\varphi_{cl}$,
$\bar{\varphi}_{cl}$, $\eta_{cl}$ and 
time reflection
$$
\varphi_{cl}(\boldsymbol{x},\, \tau) \to
  \bar{\varphi}_{cl}(\boldsymbol{x},\, -\tau)\;, \qquad
  \bar{\varphi}_{cl}(\boldsymbol{x},\, \tau) \to
  \varphi_{cl}(\boldsymbol{x},\, -\tau)\;, \qquad  
  \eta_{cl} \to \eta_{cl}\;.
$$
In general case this gives four equally suppressed saddle-point
solutions, each producing a term with complex exponent in
Eq.~(\ref{eq:41}). We expect, however, to find a single dominant saddle-point
configuration, just like for other tunneling processes. Then this 
saddle point is symmetric,  
\begin{equation}
  \label{eq:37}
  \{ \varphi_{cl}(\boldsymbol{x},\, \tau),\,
  \bar{\varphi}_{cl}(\boldsymbol{x},\, \tau),\, \eta_{cl} \} \in \mathbb{R}
  \;,\qquad 
  \varphi_{cl}(\tau,\ \boldsymbol{x}) = \bar{\varphi}_{cl}(-\tau,\,
  \boldsymbol{x}) \qquad \mbox{at}
  \;\;\;\;\tau\in \mathbb{R}\;.
\end{equation}
Relations~(\ref{eq:37})  turn quasiperiodicity~(\ref{eq:40})
into the boundary conditions (\ref{eq:44}), (\ref{eq:43}) at $\tau=-\beta/2$
and $\tau = 0$, where 
\begin{equation}
  \label{eq:23}
  \eta_0 \equiv \eta_{cl} - \omega \beta
\end{equation}
parametrizes the solutions from now on. Notably,
Eq.~(\ref{eq:37}) implies that $\varphi_{cl}$ and $\bar{\varphi}_{cl}$
are complex conjugate to each other on the Minkowski time axis. At
real $t\equiv -i\tau$  
they describe classical evolution of waves/particles 
in the final state. Simultaneously, reality of $\varphi_{cl}$ and
$\bar{\varphi}_{cl}$ in Euclidean time ensures that the semiclassical
expression (\ref{eq:41}) for the probability is real.

Third, let us simplify the leading suppression exponent $F_Q$ in
Eq.~(\ref{eq:41}). It is natural to use real Euclidean action $S_E \equiv
-iS$, Eq.~(\ref{eq:S_E}). The two other terms in the leading exponent
can be collected into 
the action of the stationary $Q$-ball, Eq.~(\ref{eq:15}), 
\begin{equation}
  \label{eq:24}
  S_E[\varphi_Q,\, \bar{\varphi}_Q] = \beta E_Q - \omega \beta Q\;,
\end{equation}
where Eqs.~(\ref{eq:21}), (\ref{eq:S_E}), (\ref{eq:17}) were used. We
arrive at the convenient expression (\ref{eq:25}) for $F_Q$.

Fourth, it is natural to expect that the semiclassical solution is
spherically-symmetric like the $Q$-ball, i.e.\ $\varphi_{cl}$ and
$\bar{\varphi}_{cl}$ depend on $r\equiv |\boldsymbol{x}|$ and $\tau$. 

Finally, we derive the Legendre transformation formula that was used
in Sec.~\ref{sec:semicl-meth-its}. To this end we explicitly  
differentiate the suppression exponent (\ref{eq:25}) with respect
to $Q$ at fixed $\beta$, use Eqs.~(\ref{eq:17}), (\ref{eq:23}),
(\ref{eq:24}) and the property of $Q$-ball $dE_Q/dQ =
\omega$~\cite{Friedberg:1976me}. We obtain $dF_Q  / dQ =
\eta_0$ which gives physical interpretation of $\eta_0$. This relation can
be also used as a cross-check of numerical calculations. 

\subsection{Expression for the prefactor}
\label{sec:expression-prefactor}
Let us further simplify Eq.~(\ref{eq:41}). To cancel the unknown
constant $N$, we divide this expression by unity~(\ref{eq:9}) which is 
given by the same path integral as in Eq.~(\ref{eq:11}) but  
with an important distinction: now the integration runs over all
final-state configurations $\varphi(\boldsymbol{x},\, t_0)$,
$\bar{\varphi}(\boldsymbol{x},\, t_0)$, not just the ones from the
vacuum sector. As a consequence,
the dominant saddle-point configuration for that integral is the
$Q$-ball $\{ \varphi_Q,\, \bar{\varphi}_Q,\, \eta_{0} = 0
\}$, Eq.~(\ref{eq:15}). We obtain the saddle-point result similar
to Eq.~(\ref{eq:41}),
\begin{equation}
\label{eq:45}
        1 = N \cdot  \frac{\mathrm{det}^{-1/2}\hat{\cal
            D}_Q}{\sqrt{dQ/d\eta}} \;,
\end{equation}
where $\hat{\cal D}_Q$ is the same fluctuation operator as in
Eq.~(\ref{eq:41}) but in the $Q$-ball background. Using
Eq.~(\ref{eq:23}) one finds that $dQ/d\eta \to \beta^{-1}dQ/d\omega +
O(\beta^{-2})$. This factor cancels in the ratio of Eqs.~(\ref{eq:41})
and~(\ref{eq:45}),
\begin{equation}
  \label{eq:48}
        {\cal P} = \mathrm{e}^{-F_Q} \cdot
        \left(\frac{\mathrm{det}\,\hat{\cal D}_{cl}}{\mathrm{det}\,
          \hat{\cal D}_Q}\right)^{-1/2}\;,
\end{equation}
where $F_Q$ is the suppression exponent from Eq.~(\ref{eq:41}).
\begin{sloppy}

The operator $\hat{\cal D}_{cl}$ introduced in Eq.~(\ref{eq:42}) is
Hermitean; recall that it acts on perturbations ${\psi \equiv(\delta
  \varphi,\, \delta \bar{\varphi})^T}$ with quasiperiodic boundary
conditions
\begin{equation}
  \label{eq:29}
  \delta \varphi(\boldsymbol{x},\, \tau + \beta) =
  \mathrm{e}^{\eta_0} \delta \varphi(\boldsymbol{x},\, \tau)\;,
  \qquad 
  \delta \bar{\varphi}(\boldsymbol{x},\, \tau + \beta) =
  \mathrm{e}^{-\eta_0} \delta \bar{\varphi}(\boldsymbol{x},\, \tau)\;,
\end{equation}
cf.\ Eqs.~(\ref{eq:12}) and~\eqref{eq:27}.  As a consequence, the 
eigenmodes $\psi_k(\boldsymbol{x},\, \tau)$ of this operator form
orthonormal basis\footnote{We still assume that the finite-size
  spatial box is introduced.} and its eigenvalues $\lambda_k$ are
real,
\begin{equation}
  \label{eq:49}
  \hat{\cal D}_{cl}\psi_k = \lambda_k \psi_k\;, \qquad
  \int_{-\beta/2}^{\beta/2} d\tau d^3 \boldsymbol{x}  \; \psi_k^\dag\,
  \psi_{k'} = \delta_{kk'}\;, \qquad 
  \mathrm{det}\, \hat{\cal D}_{cl} = \prod_k \lambda_k\;.
\end{equation}
We immediately run into a problem: $\det \hat{\cal D}_{cl} = 0$ due
to zero eigenmodes
\begin{equation}
  \label{eq:50}
 \psi_0^{(Q)}  = B_Q^{-1/2}\begin{pmatrix}\phantom{-}\chi_{cl} \\ 
   -\bar{\chi}_{cl} \end{pmatrix} \;,\;\;\;\;
  \psi_0^{(0)}  = B_0^{-1/2}\begin{pmatrix}\partial_\tau \chi_{cl} \\ \partial_\tau
    \bar{\chi}_{cl} \end{pmatrix} \;, \;\;\;\;
    \psi_0^{(i)}  = B_i^{-1/2}\begin{pmatrix}\partial_i
      \chi_{cl} \\ \partial_i 
    \bar{\chi}_{cl} \end{pmatrix}\;,
\end{equation}
satisfying ~$\hat{\cal D}_{cl}\, \psi_0^{(\alpha)} = 0$,
where we introduced fields with finite asyptotics ${\chi_{cl} \equiv
\mathrm{e}^{-\omega \tau}\, \varphi_{cl}}$, $\bar{\chi}_{cl} \equiv
\mathrm{e}^{\omega \tau}\,  \bar{\varphi}_{cl}$ and 
normalization constants 
$B_\alpha$. Indeed, the perturbations~(\ref{eq:50}) generate phase
rotations $\varphi\to \mathrm{e}^{-i\alpha} \varphi$, time and space
shifts of the background solution via
Eq.~\eqref{eq:27}. They do not
change the integrand in Eq.~(\ref{eq:11}). 

\end{sloppy}
It is clear that the functional integration in the directions of zero
modes cannot be performed using the saddle-point method. Indeed, in the 
standard case one represents 
$$
\begin{pmatrix}
    \delta \varphi \\ \delta\bar{\varphi}
  \end{pmatrix} =
  \sum_k c_k \psi_k(\boldsymbol{x},\,\tau) \;, \qquad \qquad {\cal
    D}\delta\varphi\, {\cal D}\delta\bar{\varphi} = \prod_k
  \frac{dc_k}{\sqrt{2\pi}}
$$
and takes the Gaussian integrals over $c_k$. On the other hand,
parameter $c_0$ in front of $\psi_0^{(0)}(\boldsymbol{x},\, \tau)$
corresponds to time shift  
$\delta t = -i c_0 B_0^{-1/2}$. The respective integral is
not  Gaussian; rather, it gives the full period of the process, 
\begin{equation}
  \label{eq:56}
  \int \frac{dc_0}{\sqrt{2\pi}} = i t_0\left(
  \frac{B_0}{2\pi}\right)^{1/2} \;.
\end{equation}
Thus, we should exclude zero eigenvalue of $\psi_0^{(0)}$ from the product in
Eq.~(\ref{eq:49}) and use Eq.~(\ref{eq:56}) instead. Dividing both
sides of Eq.~(\ref{eq:48}) by $t_0$, we obtain expression for the the
rate $\Gamma_Q \equiv {\cal P}/t_0$. 

The modes due to phase rotations and spatial translations are
treated in the same way as $\psi_0^{(0)}$. But in contrast to
$\psi_0^{(0)}$, they  exist in the background of $Q$-ball;
the respective expressions are obtained by changing $\chi_{cl},\,
  \bar\chi_{cl}$ to $\chi_Q(r)$ in Eq.~(\ref{eq:50}). As a
consequence, the contributions of these modes partially cancel in
Eq.~(\ref{eq:48}) leaving the ratios of the normalization 
constants $B_\alpha^{1/2}$ computed for the semiclassical solution and
$Q$-ball, cf.\ Eq.~\eqref{eq:56}. Explicit calculation gives,  
$$
  B_Q \to  \cosh\eta_0 \cdot B_Q\big|_{\chi_Q} \;, \qquad 
  B_i \to  \cosh\eta_0 \cdot B_i\big|_{\chi_Q} \qquad \mbox{as}
  \;\;\;\; \beta \to +\infty\;,
$$
where the asymptotics (\ref{eq:26}) were used in Eq.~(\ref{eq:49}).

Collecting all factors, we obtain expression \eqref{eq:57} for
the prefactor in Eq.~(\ref{eq:8}), where $\det'$ is the product
of all nonzero eigenvalues of the operator and we computed the norm $B_0$
of $\psi_0^{(0)}$ using Eq.~(\ref{eq:49}). By definition,
the determinants in Eq.~\eqref{eq:57} involve quasiperiodic 
boundary conditions~\eqref{eq:29} with $\beta \to +\infty$. However,
once all zero modes are excluded, any kind of 
boundary conditions with finite $\delta \varphi$ can be imposed at $\tau \to \pm 
\infty$. Indeed, the eigenvalue problem for $\hat{\cal
  D}_{cl}$ has the form of a  stationary Schr\"odinger equation for
the four-dimensional particle with a spin. The regions $\tau\to \pm \infty$
are classically forbidden for this particle: the 
background solution at large $|\tau|$ coincides with the $Q$-ball
which is classically stable, i.e.\ has only exponentially growing or
decaying with $\tau$ perturbations $\psi_k$. It is clear that the
values of $\lambda_k$ do not depend on the boundary conditions
imposed deep inside the classically forbidden regions. Thus, 
one can consider all determinants in Eq.~\eqref{eq:57} with vanishing
boundary conditions  ${\delta \varphi,\, \delta \bar{\varphi} \to 0}$
as ${\tau \to \pm\infty}$ instead of the quasiperiodic ones.  
 
\subsection{Finite-temperature solutions}
\label{sec:decay-at-finite}
Now, consider  $Q$-ball decay at a finite temperature $T\equiv
\beta^{-1}$. The initial state of this process is specified by the
density matrix $\hat{\rho}_{\beta} = Z_\beta \, \mathrm{e}^{-\beta
  \hat{H}}\hat{P}_Q$, where $\hat{P}_{Q}$ restricts the statistical
ensemble to states with charge $Q$, while  $Z_\beta$ is a
normalization constant ensuring  ${\mathrm{tr}\, \hat{\rho}_\beta =
1}$. As before, we tacitly assume that $\hat{\rho}_\beta$ is projected onto the
states $|i\rangle$ from the $Q$-ball sector. Note, however, that the
quantity~(\ref{eq:3}) used in the 
above calculation involves the same density matrix $\hat{\rho}_\beta' 
\equiv \mathrm{e}^{\beta    (E_Q - \hat{H})} \hat{P}_Q$ but with
different normalization. Moreover, Eq.~(\ref{eq:9}) at
finite $\beta$ gives,
$$
\mathrm{e}^{\beta E_Q} \sum_{i,\, \Psi} \;\left|   \langle \Psi|
  \mathrm{e}^{-i\hat{H}(t_0 - i\beta/2)  } \hat{P}_Q |
  i\rangle\right|^2 =  \mathrm{tr}_i \,\rho_\beta'\,
$$
  where the trace is taken in the  $Q$-ball sector. Thus, the ratio of
Eqs.~(\ref{eq:3}) and (\ref{eq:9}) at finite $\beta$ that was computed
above gives the probability
of $Q$-ball decay at temperature~$T=\beta^{-1}$.  

We conclude that the exponent $F_{Q,\, \beta}$ of thermal decay is
given by Eq.~(\ref{eq:25}), but with finite time interval $-\beta/2 <
\tau < \beta/2$ in the Euclidean actions~\eqref{eq:S_E}, while the finite-$T$ 
saddle-point solutions $\varphi_{cl}$, $\bar{\varphi}_{cl}$
satisfy Eqs.~(\ref{eq:44}) at $\tau = -\beta/2$. 

To generalize Eq.~\eqref{eq:57} for the prefactor, one should
avoid simplifications related to the limit $\beta\to +\infty$. At
finite temperature the quantities $dQ/d\eta_{cl}$ in
Eqs.~(\ref{eq:41}) and (\ref{eq:45}) do not cancel, the original
quasiperiodic conditions~\eqref{eq:29} should be used for all
operators, and orthogonal family of zero modes with norms $B_{\alpha}$
should be properly defined. We leave derivation of finite-temperature
prefactor to interested readers. 

\section{Numerical implementation}
\label{sec:numer-impl}
In this Section we solve the boundary value
problem for the semiclassical solution $\varphi_{cl}(r,\, \tau)$,
$\bar{\varphi}_{cl}(r,\,\tau)$. To this end we adopt units with $m=1$ and set
$v=1$ by field rescaling.

\subsection{Framework}
\label{sec:framework}
It is convenient to consider functions with finite asymptotics,
\begin{equation}
  \label{eq:46}
  y_{cl}(r,\, \tau) = r\, \mathrm{e}^{-\omega\tau}\, \varphi_{cl} \;,
  \qquad   \bar{y}_{cl}(r,\, \tau) = r \,
  \mathrm{e}^{\omega\tau}\, \bar{\varphi}_{cl} \;,
\end{equation} 
where $\omega$ is the frequency of the decaying $Q$-ball. We
introduce\footnote{We take larger $N_\tau$ for solutions with large periods $\beta$.}
$N_r \times N_\tau = 150\times 300$ lattice with sites
$\{r_j,\, \tau_i\} = \{ \Delta r (j+1),\, -\Delta \tau (i+1/2)\}$, $0
\leq j \leq N_r$, $0 \leq i \leq N_\tau$, and 
uniform spacings $\Delta r$,  $\Delta \tau$. The spatial and temporal
extents of our lattice $r_j < L_r$ and $-\beta/2 <  \tau_i <
0$  are large enough for the solutions to approach their
asymptotics: $L_r 
= 20$, $\beta = 40\div 200$. Our unknowns are 
the values of $y_{j,\, i} \equiv y_{cl}(r_j ,\, \tau_i)$ and 
$\bar{y}_{j,\,i}$ on the lattice sites.

We discretize the field equations~(\ref{eq:6}) in the standard
second-order manner using
$$
\partial_\tau y_{cl}(r_j,\,\tau_i) \to  \frac{y_{j,\, i-1} - y_{j,\,
      i+1}}{2\Delta \tau}\;, \qquad 
  \partial_\tau^2 y_{cl}(r_j,\,\tau_i) \to  \frac{y_{j,\, i-1}
    + y_{j, \, i+1}- 2y_{j,\, i}}{\Delta \tau^2}
$$
and similar expressions for the derivatives with respect to
$r$. Boundary conditions  (\ref{eq:44}), (\ref{eq:43}) have natural
lattice generalization,
\begin{align}
  \label{eq:36}
  & y_{j,\, N_\tau} = \bar{y}_{j,N_{\tau}-1}\;\mathrm{e}^{-\eta_0}  &&
  \bar{y}_{j,\, N_\tau} = y_{j,N_{\tau}-1}\;\mathrm{e}^{\eta_0}
  \\\notag
  & y_{j,-1} = \bar{y}_{j,0} \;, && \bar{y}_{j,\, -1} = y_{j,0} \;,
\end{align}
where $\tau_0 = -\Delta \tau/2$ and $\tau_{N_{\tau}-1} = -\beta/2 +
\Delta \tau /2$ are the first and the last lattice sites. Finally, we
impose Dirichlet and Neumann conditions at $r = 0$ and $L_r$,
respectively,
\begin{equation}
  \label{eq:39}
  y_{ -1,\, i} = \bar{y}_{-1,\, i} = 0\;, \qquad y_{ N_r,\, i} = y_{
    N_r - 1,\, i} \;, \qquad \bar{y}_{N_r,\, i} = \bar{y}_{
    N_r - 1,\, i} \;,
\end{equation}
cf.\ Eq.~(\ref{eq:46}). After discretization Eqs.~(\ref{eq:6}) with
boundary conditions (\ref{eq:36}), (\ref{eq:39}) constitute a sparse
system of $2N_rN_\tau$ nonlinear algebraic equations for the
same number of unknowns $\varphi_{j,\, i}$, $\bar{\varphi}_{j,\, i}$.

A convenient technique to solve large systems of this kind
is provided by the Newton-Raphson method,  see e.g.\ \cite{Press}. One 
starts with the initial guess $y^{(0)}_{j,\, i}$, $\bar{y}^{(0)}_{j,\,
  i}$ for the solution. Substituting $y = y^{(0)} + \delta y$,
$\bar{y} = \bar{y}^{(0)} + \delta \bar{y}$ into the lattice equations and
ignoring nonlinear terms in $\delta y_{j,\,   i}$, $\delta
\bar{y}_{j,\,   i}$, one arrives at the sparse linear system that
gives $\delta y$ and $\delta\bar{y}$. Then one
refines the guess by setting $y^{(0)} \to 
y^{(0)}  + \delta y$, $\bar{y}^{(0)} \to y^{(0)} + \delta \bar{y}$ and repeats
the procedure. The iterations converge to the exact solution if the
original guess was good enough. Finally, one adjusts the value of
$\eta_0$ in Eq.~(\ref{eq:36}) to fix the charge of the solution to
that of the $Q$-ball with frequency $\omega$.  Note that the above
algorithm should be supplemented with fast linear solver, cf.~\cite{Demidov:2015nea}.

\subsection{Obtaining the solutions}
\label{sec:starting-configuration}
A drawback of the Newton-Raphson method is its high sensitivity
to the initial guess: if $y^{(0)}$, $\bar{y}^{(0)}$ are not close
enough to the semiclassical solution, the method may diverge or produce useless
trivial solutions~--- the $Q$-ball, $Q$-cloud, or charged condensate
in the vacuum. This  poses a problem of finding an approximate
solution at some $\beta$ and~$\eta_0$.  
\begin{sloppy}

To construct such a solution, we use a nontrivial
observation~\cite{Khlebnikov:1991th} that the (quasi)periodic
finite-$\beta$ solutions describe, besides the thermal decays, transitions
between the sectors of $Q$-ball and vacuum at fixed energies
$E$. Indeed, thermal density matrix is diagonal in the basis of
energy eigenstates, which implies that the semiclassical solutions
describing fixed-energy processes contribute with factors
$\mathrm{e}^{-\beta   E}$ to the thermal rates. This gives physical
interpretation to all our quasiperiodic solutions, even the ones  
producing subleading contributions in  
Fig.~\ref{fig:Fbeta}b.

\end{sloppy}
A particular semiclassical solution with energy $E$ close to the
$Q$-cloud mass $\tilde{E}_Q$ describes transmission through the
infinitesimally small classically forbidden region near the barrier
top in Fig.~\ref{fig:EQ2}b. It should coincide with the $Q$-cloud, a
configuration ``sitting'' on top of the barrier, up to small
corrections. We  demonstrated in Appendix
\ref{sec:appendix-solitons}, however, that almost all small 
perturbations of the $Q$-cloud grow exponentially with $\tau$ and 
therefore cannot satisfy the quasiperiodic conditions
(\ref{eq:44}). The notable exception is the negative (unstable) mode
which periodically depends on $\tau$ via  $\mathrm{e}^{i\gamma \tau}$. In
Fig.~\ref{fig:E-T-Q353}a we plot the profile $\xi_- (r)$ of this mode
obtained in Appendix~\ref{sec:appendix-solitons}.  
\begin{figure}[htb]

 \vspace{.9cm}
  
\hspace{3.2cm}(a) \hspace{7.5cm}(b)

\vspace{-.9cm}
%
  \centerline{
    \includegraphics[scale=0.973]{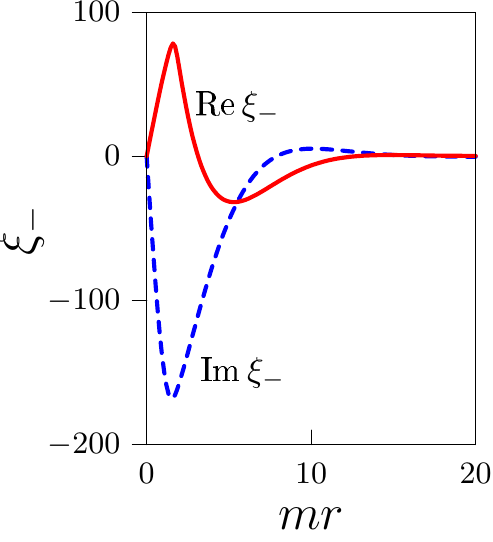}\hspace{10mm}
    \includegraphics{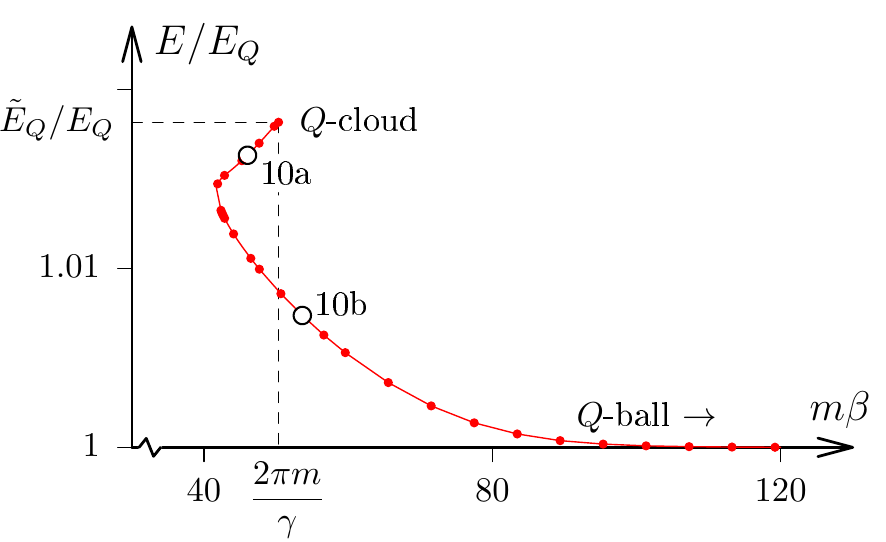}}
%
\caption{(a) Negative mode of the $Q$-cloud $\xi_-(r)$ at $Q\approx 1.33\,
  Q_c$. (b) Energies $E$ of the semiclassical solutions with different 
  Euclidean periods $\beta$ and the same charge $Q \approx 1.33\, Q_c$
  (small filled points). Empty circles with numbers represent
  solutions in Fig.~\ref{fig:solT}, the solution in
  Fig.~\ref{fig:solutions}b describing $Q$-ball decay
   is obtained in 
  the limit~$\beta \to +\infty$. \label{fig:E-T-Q353}}
\end{figure}
Adding it  to the $Q$-cloud $\tilde{y}_Q(r)$ 
we obtain the  approximate solution with charge $Q$ and period $\beta =
2\pi/\gamma$, 
\begin{align}
\label{eq:starting-configuration}
& y_{cl} = \left[\tilde{y}_{Q} + A_- \,(\xi_- e^{-i\gamma\tau} + \xi_-^*
e^{i\gamma\tau})\right] \mathrm{e}^{\tau (\tilde{\omega} - \omega)}\;,\\  \notag
& \bar{y}_{cl} = \left[\tilde{y}_{Q} + A_-(\xi_- e^{i\gamma\tau} + \xi_-^*
e^{-i\gamma\tau}) \right]\mathrm{e}^{-\tau (\tilde{\omega} - \omega)}\;.
\end{align}
where we used Eq.~(\ref{eq:34}) from Appendix
\ref{sec:appendix-solitons} and recalled that $\xi_-^*(r) \,
\mathrm{e}^{-i\gamma\tau}$ also satisfies the linearized field
equations in the background of $Q$-cloud. Besides,  we introduced 
the $Q$-cloud frequency $\tilde{\omega}$ and small amplitude $A_-$ of
the perturbation. It is straightforward to check that the configuration 
(\ref{eq:starting-configuration}) is real and satisfies the 
semiclassical equations (\ref{eq:6}), (\ref{eq:44}), (\ref{eq:43})  
with $\eta_0 = \beta (\tilde{\omega} - \omega)$. 

\begin{figure}[htb]
  \centerline{
    \includegraphics{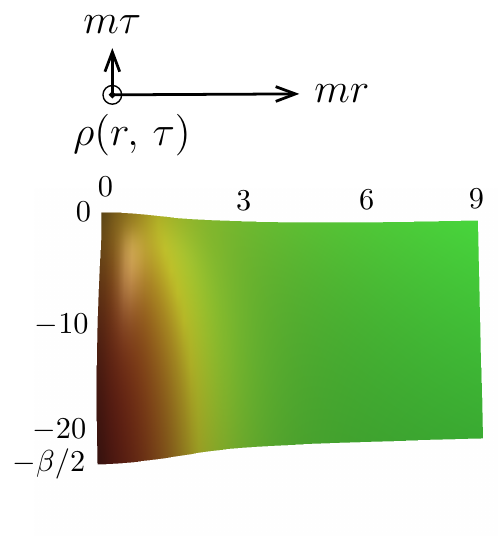}\hspace{12mm}
    \includegraphics{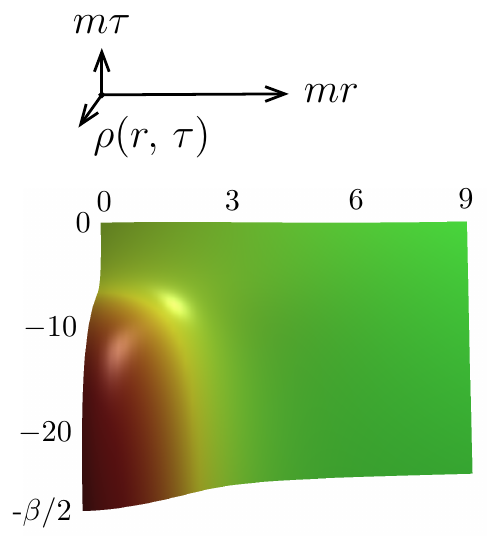}\hspace{12mm}
    \includegraphics{fig_legend.pdf}}
   %

  \vspace{-4mm}
  \hspace{3.3cm}(a) \hspace{6.0cm}(b) 
    \caption{Numerical solutions $\rho = (\varphi
      \bar{\varphi})^{1/2}$ with $Q\approx 1.33\, Q_c$ and different
      Euclidean periods: (a)~$m\beta \approx 46$, (b) $m\beta \approx
      54$. \label{fig:solT}}
\end{figure}

This suggests the following strategy of numerical calculations at 
a given charge~--- say, at $Q \approx 1.33\, Q_c$. We compute 
the configuration~(\ref{eq:starting-configuration}) with small $A_- =
10^{-3}$  and use it as an initial guess for the solution with
period ${\beta= 2\pi / \gamma - \delta \beta}$, $\delta \beta = 0.2$
and $\eta_0 = 5.7$. After several Newton-Raphson iterations the new solution is
obtained with acceptable precision. Then we use this 
solution, in turn, as an initial guess at even
smaller period 
$\beta$. Continuing  to change the value of $\beta$ in small steps, we   
obtain a complete\footnote{Note that the Newton-Raphson method
  diverges near the leftmost point of the plot in
  Fig.~\ref{fig:E-T-Q353}b where two almost identical solutions with 
  equal periods exist.  This restricts the above procedure of small
  deformations to the upper branch in
  Fig.~\ref{fig:E-T-Q353}. Nevertheless, we have found the solutions
  from the lower branch by using the initial
  guess~(\ref{eq:starting-configuration}) at $\beta > 2\pi / \gamma$, 
  beyond the region of its applicability. We did not need the
  regular procedure for passing the turn-arounds which prescribes to
  add Eq.~(\ref{eq:21}) to   
  the system of lattice equations and parametrize the solutions with
  energy $E$  instead of~$\beta$.} branch of    
solutions shown by points in  Fig.~\ref{fig:E-T-Q353}b, see examples in
Fig.~\ref{fig:solT}. At last, we
arrive at the solution with large $\beta$ (Fig.~\ref{fig:solutions}b)
describing decay of an isolated $Q$-ball. The suppression exponents 
(\ref{eq:25}) of our numerical solutions are shown in~Fig.~\ref{fig:Fbeta}b. 

Once the solution with large $\beta$ is found, we start changing
$\omega$ and $Q(\omega)$ in small steps. As before, at each step we
use the solution with charge $Q$ as an initial guess for the one with  
charge $Q\pm \delta Q$. This gives the semiclassical solutions in the
entire metastability region $Q_c < Q < Q_s$. We compute their
suppression exponents (\ref{eq:25}) and obtain Figs.~\ref{fig:result}b
and \ref{fig:Fbeta}a. 

We supported the above numerical procedure with several 
tests. First, we checked that all numerical solutions with large
$\beta$ are close to the $Q$-balls at $\tau \approx -\beta/2$, see
Fig.~\ref{fig:sections}a, while their energies $E = E(Q)$ coincide
with $E_Q$. Second, the suppression exponent $F_Q$ computed
via Eq.~(\ref{eq:25}) has the expected properties: it equals zero at $Q =
Q_c$ and grows to infinity as $Q  \to Q_s$.  Third, the numerical values
of $F_Q$ and $\eta_0$  are consistent with the Legendre 
relation $dF_Q/dQ = \eta_0$. Fourth, we checked that the semiclassical
solutions and the values of $F_Q$ are not sensitive to the lattice
parameters $L_r$, $\Delta r$, $\Delta \tau$ within relative precision
of $2\%$. Besides, we monitored the energy $E$ and 
charge $Q$ of the solutions which were conserved  with accuracy
better than $10^{-3}$. These tests convinced us that the lattice solutions
give correct suppression exponent at the precision level of~$2\%$.

\section{Discussion: Prospects for cosmology}
\label{sec:gener-prosp}
We developed general semiclassical method to calculate the
decay rate ${\Gamma_Q = A_Q \cdot \mathrm{e}^{-F_Q}}$ of metastable  
$Q$-balls at $Q\gg 1$. The method can be applied in arbitrary models
at the cost of numerically obtaining certain Euclidean solutions
$\varphi_{cl}(\boldsymbol{x},\, \tau)$ that enter the
semiclassical expressions for exponent $F_Q$ and 
prefactor $A_Q$ of the rate. We generalized the method to
finite-temperature processes.

\begin{figure}[htb]

  \vspace{3mm}
  \hspace{4cm} (a) \hspace{6.5cm} (b)

  \vspace{-3mm}
  \centerline{
    \includegraphics{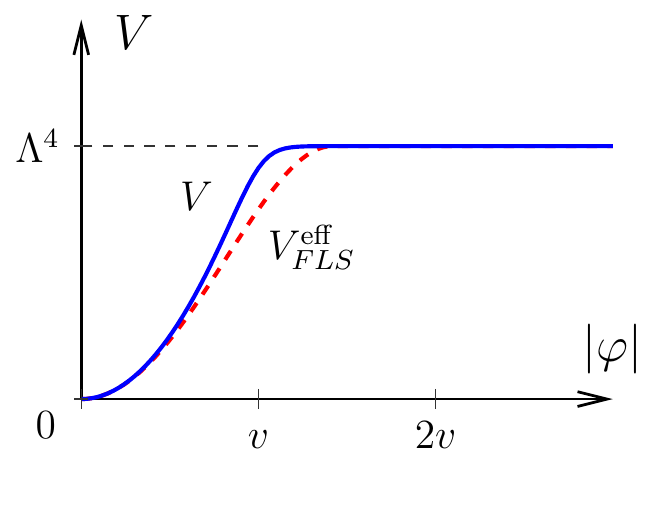}\hspace{5mm}\includegraphics{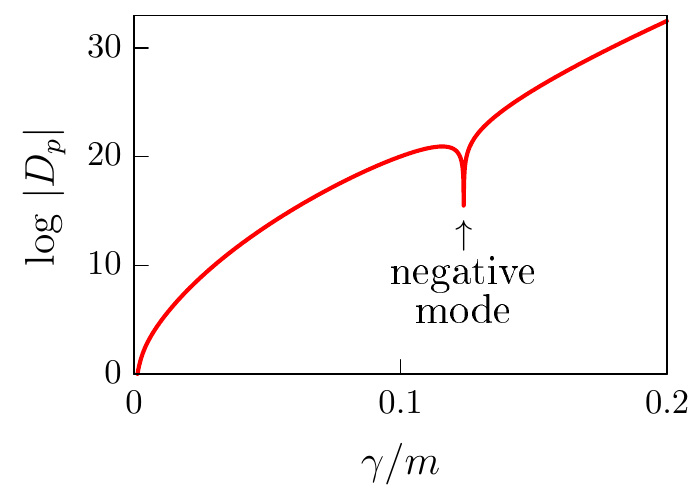}}

  \caption{(a) Our potential (\ref{eq:2}) compared  to the effective
    low-energy potential~(\ref{eq:19}) of the FLS
    model (solid and dashed lines, respectively). (b) Determinant
    $D_{p}$ of the matrix in Eq.~(\ref{eq:38}) as a function of
    the growth rate~$\gamma$. The graph is obtained for the $Q$-cloud
    with charge $Q\approx 
    1.33 \, Q_c$. \label{fig:FLS}}
\end{figure}

To illustrate the method, we  performed explicit
numerical calculations in the model~(\ref{eq:2}). In particular, we computed the
exponent $F_Q$  of the $Q$-ball decay rate, see Eq.~(\ref{eq:32}), and
estimated\footnote{Note that 
  numerical evaluation of the fluctuation determinant in the prefactor
  involves application of two-dimensional Gelfand-Yaglom theorem and
  renormalization, which are interesting open tasks beyond the scope
  of this paper, see~\cite{Callan:1977pt, Baacke:2003uw,
    Dunne:2005rt,Andreassen:2016cvx}.} the prefactor $A_Q \sim
mQ_c^{1/2}$. Notably, the model
we use is close to a certain limit of the 
celebrated Friedberg-Lee-Sirlin (FLS) model~\cite{Friedberg:1976me}
which describes complex and real fields $\varphi$ and $\zeta$ with
renormalizable potential  
\begin{equation}
  \label{eq:18}
  V_{FLS} = (\lambda \zeta^2 - \Lambda^2)^2 + \frac{\Lambda^2}{v^2} \,
  \lambda \zeta^2 |\varphi|^2\;.
\end{equation}
Here $m \equiv \Lambda^2 / v$ and $m_\zeta \propto \lambda^{1/2}
\Lambda$ are masses of the fields, $\lambda$ is the coupling
constant. At $m_\zeta \gg m$ the field $\zeta$ is very massive and
cannot be excited in processes with relatively low momenta  $p \sim 
m$. Minimizing Eq.~(\ref{eq:18}) with respect to $\zeta$ at fixed
$\varphi$, one obtains low-energy potential for the remaining field,
\begin{equation}
  \label{eq:19}
  V_{FLS}^{\mathrm{eff}} = \Lambda^4 - \Lambda^4 \left( 1 -
  |\varphi|^2/2v^2 \right)^2 \theta (2v^2 - |\varphi|^2)\;.
\end{equation}
This potential is close to the smooth one, Eq.~(\ref{eq:2}) that was
used in our calculations, see Fig.~\ref{fig:FLS}a. Thus, our numerical
data estimate decay rate of the FLS $Q$-balls.

Although we did not directly address supersymmetric models with flat
directions~\cite{Enqvist:2003gh}, the decay rate of their $Q$-balls 
is still expected to exhibit the behavior similar to that in 
Fig.~\ref{fig:result}b.  In particular, $\Gamma_Q$ should be
unsuppressed at the critical charge $Q_c$ corresponding to classically
unstable $Q$-ball, it should sharply decrease with charge at $Q>Q_c$
and reach zero at the point of absolute stability $Q=Q_s$. Deep inside
the metastability window $Q_c  < Q <  Q_s$  the suppression exponent
of the rate should be large, $F_Q\sim Q_c\gg 1$.

Let us discuss possible cosmological application of  metastable
$Q$-balls with $Q\sim Q_c$. These objects may form dark matter with
naturally large lifetime, so that their decays in the late-time
Universe produce warm dark matter component~--- energetic
$\varphi$-particles leaving the galaxies.  The latter process may 
change structure formation or explain~\cite{Chudaykin:2016yfk}
appearing  tension between the low-redshift~\cite{Riess:2011yx,
  Freedman:2012ny} and CMB~\cite{Ade:2015xua} measurements of the
Hubble constant.

Note that the decays of dark matter $Q$-balls are observable only if
their lifetime $\Gamma_Q^{-1}$ is comparable to the age of the Universe
or  smaller. This constrains the $Q$-ball charge $Q\sim Q_c
\lesssim 150$, where we used estimates ${F_Q  \sim Q_c}$, ${A_Q
  \sim m Q_c^{1/2}}$ in the mid of the metastability window and 
assumed wide interval of masses ${m=1 \,   \mbox{eV}   \div 10^{19}
  \,   \mbox{GeV}}$ in the prefactor. Thus, we need a mechanism to
generate small $Q$-balls in the early Universe. 

The two standard mechanisms of forming dark matter $Q$-balls involve
phase transition~\cite{Frieman:1988ut, Griest:1989cb, Krylov:2013qe} 
or fragmentation of the scalar condensate~\cite{Kusenko:1997si,
  Kasuya:2014bxa, Zhou:2015yfa} at high temperature
$T$. Typically, these mechanisms act in a similar way, by collecting 
charge $Q$ from the cosmological horizon of size $H^{-1}$ into the compact objects
which further relax into   $Q$-balls. Assuming  initial charge
asymmetry\footnote{Possibly   generated via the Affleck-Dine
  mechanism~\cite{Affleck:1984fy, Dine:2003ax, 
    Allahverdi:2012ju}.} $\Delta_Q = n_Q/s$, where $n_Q$ 
and $s$ are the charge and entropy densities at temperature $T$, one
estimates the charges of formed $Q$-balls,
\begin{equation}
  \label{eq:16}
  Q \approx \frac{4\pi}3 \, \Delta_Q s\,  H^{-3} \sim
  \frac{\Delta_Q M_{pl}^3}{ T^3\sqrt{g_*} } \sim 10^2\;,
\end{equation}
where in the second equality we introduced the number of relativistic
degrees of freedom  $g_* \sim 10^2$, used Friedmann equation $H
\approx g_*^{1/2} T^2/M_{pl}$ and expression for the entropy density $s
\sim g_* T^3$. After that we recalled that $Q\sim
10^2$. Equation (\ref{eq:16}) relates the generation temperature $T$ 
to the asymmetry $\Delta_Q$. 

The present-day mass density of $Q$-balls should coincide
with that of dark matter $\rho_{DM} \sim 10^{-6} \,
\mbox{GeV}/\mbox{cm}^3$. Immediately after generation the
concentration of these objects is $n_{Q\mathrm{-balls}} = n_Q/Q  = s
\Delta_Q/Q$. Since expansion of the Universe conserves the ratio 
$n_{Q\mathrm{-balls}}/s$, the mass density of
$Q$-balls at the present epoch is 
\begin{equation}
  \label{eq:20}
  \frac{E_Q}{Q} \, s_0 \Delta_Q \approx 
  m s_0 \Delta_Q \sim \rho_{DM}\;,
\end{equation}
where we introduced the entropy density now $s_0\sim 10^3  \,
\mbox{cm}^{-3}$ and estimated the mass of a small 
$Q$-balls as $E_Q \sim mQ$.

Equations (\ref{eq:16}) and (\ref{eq:20}) relate parameters $m$,
$T$, and $\Delta_Q$ of the cosmological model. The standard generation
mechanisms naturally work at comparable values of $m$ and $T$. Taking
$T\sim 10^{-2} m$ to prevent immediate decay of $Q$-balls due to the 
temperature fluctuations (see Fig.~\ref{fig:Fbeta}b), we obtain
$\Delta_Q \sim 10^{-22}$ and $m\sim 10^2 \,T{\sim 10^{13} \,
\mbox{GeV}}$. Note that this generation temperature is much higher than
the ones in typical models with large $Q$-balls,
cf.~\cite{Frieman:1988ut, Griest:1989cb, Krylov:2013qe,
  Kusenko:1997si, Kasuya:2014bxa, Zhou:2015yfa}. Similar small-mass
$Q$-balls were considered as natural candidates for self-interacting
dark matter~\cite{Kusenko:2001vu}.
\begin{sloppy}

Since the value of $\Delta_Q$ is way smaller than the observed baryon
asymmetry ${\Delta_B \sim 10^{-10}}$, we cannot interpret the 
charge $Q$ as the baryon number thus relating small $Q$-balls to the  
Affleck-Dine baryogenesis. Indeed, bold substitution of $\Delta_Q
=\Delta_B$ into Eqs.~(\ref{eq:16}), (\ref{eq:20}) gives essentially
different  values of mass $m\sim \mbox{GeV}$ and temperature $T\sim
10^{15} \, \mbox{GeV}$ which is problematic for the generation  
mechanisms. Hence, in order to pack the baryon number inside 
metastable $Q$-balls one has to consider special mechanisms of their
formation or at least some modification of the standard ones.
Alternatively, one can avoid identification of the charge $Q$ with
the baryon number, since cosmology of small metastable $Q$-balls is
interesting enough by itself.  

\end{sloppy}
\paragraph{Acknowledgments.}
We thank Dmitry Gorbunov and Mikhail Smolyakov for fruitful
discussions. This work was supported by the grant RSF
16-12-10494. Numerical calculations  were performed on the
Computational cluster of the Theory Division of INR RAS. 

\appendix

\section{Solitons and their (in)stability}
\label{sec:appendix-solitons}
Here we study nontopological solitons in the model
(\ref{eq:2}) and linear perturbations in their backgrounds. We use
the units with $m=1$ and set $v=1$ by field rescaling.
\begin{sloppy}

We compute the soliton profiles $\chi_Q(r)$ using the shooting
method. First, we substitute the Ansatz~(\ref{eq:15}) into the
classical field  equations and arrive to
\begin{equation}
  \label{eq:30}
  \partial_r^2 y_Q  = - \omega^2 y_Q - V'(y_Q^2/r^2)\,  y_Q\;,
  \qquad\qquad  y_Q \equiv r \chi_Q(r) \in \mathbb{R}\;,
\end{equation}
where the prime is a derivative of $V$ with respect to its argument. Second, we solve
Eq.~(\ref{eq:30}) with initial Cauchy data $y_Q(0) = 0$,
${\partial_r y_Q(0) = \chi_0}$ at $r=0$.  This can be done 
by many efficient numerical methods. It is convenient, however, to
discretize Eq.~(\ref{eq:30}) on the same spatial lattice $\{r_j \}$ as in 
Sec.~\ref{sec:numer-impl} and then sequentially determine all
$y_Q(r_j)$ using discrete evolution in Eq.~(\ref{eq:30}) from $r = 0$
to large $r = r_{N_r}$. Third, we adjust the  parameter $\chi_0$ in
such a way that the Neumann boundary condition 
$y_Q(r_{N_r-1}) = y_Q(r_{N_r})$ is satisfied at the rightmost link of 
the lattice. This gives for every~$\omega$ a real soliton
$\{y_Q(r_j)\}$  which can be conveniently used in the subsequent
lattice calculations, see Fig.~\ref{fig:Qw}a. The charge 
$Q$ and  energy  $E_Q$ of  the solution
are obtained by substituting the Ansatz (\ref{eq:15}) into
Eqs.~(\ref{eq:1}), (\ref{eq:21}) and computing the integrals over~$r$.

\end{sloppy}
Now, consider linear instabilities of nontopological solitons. 
In few models~\cite{Marques:1976ri, Gulamov:2013ema} this analysis can
be performed analytically, but we need general numerical method to
find growing  soliton modes. We add
small 
time-dependent perturbation $\xi_- \mathrm{e}^{\gamma 
t}$ to the soliton profile $y_Q(r)$,
\begin{equation}
  \label{eq:34}
  \varphi(r,\, t) = \left[ y_Q(r) + \xi_-(r)\, \mathrm{e}^{\gamma t}\right] \,
  \frac{\mathrm{e}^{i\omega t}}{r}\;, \qquad \qquad \xi_- = \xi_R + i\xi_I\;.
\end{equation}
It was shown in~\cite{Panin:2016ooo} that all growing perturbations of
this kind, if exist, have real~$\gamma$. Substituting Eq.~(\ref{eq:34}) into the
classical field equations and linearizing them with respect to small
$\xi_R$, $\xi_I$, we find,
\begin{align}
\label{eq:eta-divergent-modes-2}
&  \partial_r^2\xi_R  =  (\gamma^2 - \omega^2)\,\xi_R - 2\omega\gamma\,\xi_I + V'_Q
\, \xi_R + 2 V_Q''\, \chi_Q^2
\, \xi_R ,\\ \notag
 &\partial_r^2\xi_I = (\gamma^2 - \omega^2)\,\xi_I + 2\omega\gamma\, \xi_R +
V'_Q \, \xi_I \;,
\end{align}
where $V_Q\equiv V(\chi_Q^2)$. Note that complex conjugate
perturbation $\xi_-^* \equiv \xi_R - i\xi_I$ satisfies
Eqs.~(\ref{eq:eta-divergent-modes-2}) with the growth rate $(-\gamma)$.

We want to obtain the parameters $\gamma$ of regular perturbations
$\xi_{R}$, $\xi_I$. To this end we
appeal to the shooting method, again. We discretize 
Eqs.~(\ref{eq:eta-divergent-modes-2}) using the lattice $\{r_j \}$ from 
Sec.~\ref{sec:numer-impl} and impose Neumann condition $\xi_{R,\,
  I}(r_{N_r-1}) = \xi_{R,\, I} (r_{N_r})$ at the rightmost link of the
lattice. After
that we introduce two basis solutions $\Psi^{(A)}(r_j)  
\equiv (\xi_{R}^{(A)},\, \xi_{I}^{(A)})$ and $\Psi^{(B)}(r_j)$ satisfying, in
addition, the Dirichlet conditions $\Psi^{(A)}(r_{N_r}) = ( 1,\, 0)$ and
${\Psi^{(B)}(r_{N_r})   =  (0,\, 1)}$ at the rightmost lattice site. This gives a
full set  of Cauchy data for $\Psi^{(A)}(r)$ and $\Psi^{(B)}(r)$. Solving
Eqs.~(\ref{eq:eta-divergent-modes-2})  from $r = r_{N_r}$ to 
$r=0$, one finds these functions at all lattice sites. Now, recall 
that we are searching for a specific solution $\Psi(r_j) = c_A \Psi^{(A)}(r_j)  
+ c_B \Psi^{(B)}(r_j)$ which is regular at the origin,
i.e.\ satisfies $\Psi(0) = 0$. This gives the system of linear equations,
\begin{equation}
  \label{eq:38}
\hat{D}_{p} \begin{pmatrix} c_A \\ c_B \end{pmatrix} = 0\;, \qquad
\qquad \hat{D}_p = \begin{pmatrix} \xi_{R}^{(A)}(0) &
  \xi_{R}^{(B)}(0) \\
  \xi_{I}^{(A)}(0) &
  \xi_{I}^{(B)}(0) \end{pmatrix}\;, 
\end{equation}
where the elements of the matrix $\hat{D}_p$ representing
$\Psi^{(A)}(0)$ and $\Psi^{(B)}(0)$ 
are already known. Equation~(\ref{eq:38}) has nontrivial solutions only if $D_p
\equiv \det \hat{D}_p = 0$. In Fig.~\ref{fig:FLS}b we plot $\log
|D_p|$ as a function of $\gamma$ for the $Q$-cloud with  $\omega \approx 0.99$. 
Parameter $\gamma$ of the unstable (negative) mode corresponds to a
sharp dip in this graph, $D_p=0$. The mode $\Psi = (\xi_{R},\, \xi_I)$
in Fig.~\ref{fig:E-T-Q353}a is then obtained using Eq.~(\ref{eq:38})
and normalization condition $c_A^2 + c_B^2 = 1$.

Performing the above procedure at different $\omega$, we learn that all
$Q$-clouds in Fig.~\ref{fig:Qw}b have precisely one unstable
mode with $\gamma > 0$, while the $Q$-balls have none. At the critical point
$dQ/d\omega=0$ we obtain $\gamma=0$. 


\end{document}